\documentclass[traditabstract]{aa} 
\usepackage{txfonts}
\usepackage{natbib}
\bibpunct{(}{)}{;}{a}{}{,} % to follow the A&A style
\usepackage{graphicx}
\usepackage{color}
\usepackage{hyperref}
\usepackage{breakurl}
\usepackage{ulem}

\begin{document}

\title{Do cluster properties affect the quenching rate?}
\author{A. Raichoor\inst{\ref{Brera},\ref{GEPI}}$^,$\thanks{e-mail:\texttt{\href{mailto:anand.raichoor@obspm.fr}{anand.raichoor@obspm.fr}}} \and S. Andreon\inst{\ref{Brera}}}
\institute{
INAF -- Osservatorio Astronomico di Brera, via Brera 28, 20121 Milan, Italy\label{Brera}
\and
GEPI, Observatoire de Paris, 61, Av. de l\textquoteright Observatoire, 75014 Paris, France\label{GEPI}
}

\abstract
{The quenching rate is known to depend on galaxy stellar mass and environment, however, possible dependences on the hosting halo properties, such as mass, richness, and dynamical status, are still debated.
The determination of these dependences is hampered by systematics, induced by noisy estimates of cluster mass or by the lack of control on galaxy stellar mass, which may mask existing trends or introduce fake trends. 
We studied a sample of local clusters  (20 with $0.02 < z < 0.1$ and $\log(M_{200}/M_\odot) \gtrsim 14$), selected independent of the galaxy properties under study, having homogeneous optical photometry and X-ray estimated properties.
Using those top quality measurements of cluster mass, hence of cluster scale, richness, iron abundance, and cooling time/presence of a cool-core, we study the simultaneous dependence of quenching on these cluster properties on galaxy stellar mass $M$ and normalised cluster-centric distance $r/r_{200}$. 
We found that the quenching rate can be completely described by two variables only, galaxy stellar mass and normalised cluster-centric distance, and is independent of halo properties (mass, richness, iron abundance, presence of a cool-core, and central cooling time).
These halo properties change, in most cases, by less than 3\% the probability that a galaxy is quenched, once the mass--size ($M_{200}$ -- $r_{200}$) scaling relation is accounted for through cluster-centric distance normalisation.}

\keywords{Galaxies: clusters: general - Galaxies: evolution - Galaxies: star formation}

\titlerunning{Do the cluster properties affect the quenching rate?}
\authorrunning{A. Raichoor \& S. Andreon}
\maketitle

%@@@@@@@@@@@@@@@@@@@@@@@@@@@@@@@@@@@@@@@@@@@
% INTRODUCTION
%@@@@@@@@@@@@@@@@@@@@@@@@@@@@@@@@@@@@@@@@@@@

\section{Introduction \label{sec:intro}}

We have known for a century \citep{hubble36} that the centre of massive galaxy clusters are dominated by red early-type galaxies and that the most massive clusters are maximally dominated by this passive population.
These two trends involving cluster mass and cluster-centric radius are  informative about the effects of the environment on galaxies \citep[e.g.][]{treu03}.
At the same time, these trends are a technical obstacle to the study of the evolution of galaxies in clusters: galaxy populations cannot be compared without accounting for the (now obvious) trends with cluster-centric distance and cluster mass. 

% Cluster-centric distance
Control on the cluster-centric distance was originally implemented by \citet{butcher78} by normalising distances to a reference distance, $r_{30}$, the radius that includes 30\% of the population, to account for the different size of clusters of different mass, and thus to ``homogenise", at least in terms of cluster-centric distances, cluster samples inevitably formed by clusters of different sizes. 
Modern works kept the spirit of the original \citet{butcher78} suggestion, but now use a more physical distance, such as $r_{200}$ \citep[the radius inside which the mean density is 200 times the critical density at the cluster redshift; e.g.][]{de-propris04, andreon06, andreon06a, haines09, raichoor12, raichoor12b}.
This physical distance, being not directly observable, is derived by some observable with as little as possible scatter.
In this respect, X-ray temperature $T_X$ is preferable to a noisy velocity dispersion or to a noisy measurement of richness such as $B_{gc}$, because it possesses a low scatter with mass (see \citealt{evrard08} for velocity dispersion and \citealt{vikhlinin09} for $T_X$).
An alternative way to measure the environment is the nearest neighbour density $\Sigma_n$: though $\Sigma_n$ provides a finer estimation of the local environment, its estimation requires a careful control on possible biases and statistical analysis must account for its inherent noisiness; its use is thus less straightforward than the cluster-centric distance.

% Cluster mass
The second issue, cluster mass dependency, was raised in \citet{andreon99}, and called the ``apple vs. orange" issue or ancestor bias:  a robust determination of the galaxy evolution pattern requires that the high redshift studied clusters are the likely ancestors of the low redshift clusters in the sample. 
At the time of the \citet{butcher78} work, only ``special" clusters at high (for that time) redshift were known \citep{kron95,andreon99}.
These clusters were unlikely to be the ancestors of the low redshift clusters studied in \citet{butcher78}, being much more X-ray luminous and thus massive  than the low redshift counterparts to which they were compared \citep{andreon99}.
Assembling a sample of clusters of a fixed, or decreasing, mass with increasing redshift is far from obvious, and such samples became available when X-ray surveys were deep enough to detect and measure the mass of normal-mass clusters at intermediate-high redshift. 
\citet{andreon06} first derived for such a sample the radius-normalised quenching rate (red fraction). Note that since cluster mass grows with time, likely ancestors of today's massive clusters are clusters of intermediate mass at high redshift.
Current Sunyaev-ZelÕdovich (SZ) surveys are also providing samples of comparable mass at different redshift.

% Galaxy mass
Galaxy properties also depend on galaxy stellar mass, a point already known at least since \citet{butcher78}, who proposed the comparison of populations brighter than a fixed absolute magnitude to account for (``control" is the appropriate statistical term) this dependency.
The original \citet{butcher78} prescription has been improved over the years: \citet{andreon06} suggested the use of likely ancestors of galaxies in the  $z=0$ sample by selecting galaxies that, when evolved to $z=0$, would be part of the $z=0$ sample.
Other works proposed using galaxy stellar mass \citep[e.g.][]{de-propris03}, and moving from an integral (i.e. more massive than a given threshold)  to a differential approach \citep[i.e. per galaxy stellar mass bin; e.g. ][]{muzzin12,peng10,raichoor12,raichoor12b}.
Another improvement in this direction relates to the galaxy stellar mass threshold used to define samples at different redshifts.
Rather than fixed galaxy stellar mass, determined at the redshift of observation (as done in most works), using an evolved galaxy stellar mass \citep[e.g. given by the integral of the star formation rate over $0 \le t < \infty$, as in ][]{raichoor12,raichoor12b}, allows us to account for the galaxy stellar mass growth due to star formation episodes: in fact, a 5 $M_\odot/yr$ star-forming galaxy of $10^{10} M_\odot$ doubles its mass in  just 2 Gyr.
The failure to account for the galaxy stellar mass evolution associated with the star formation rate may introduce a spurious signal due to an unwanted drift of galaxies inside/outside the sample, a feature usually called the selection effect.

% Evolutionary cuts
To summarise, the study of galaxy evolution in clusters requires the control of at least cluster mass, cluster-centric distance (which requires knowledge of cluster mass), and galaxy stellar mass.
When galaxy populations are split in classes, we also need to know how the boundaries defining the classes should evolve with time to keep them consistently across redshifts.
The static \citet{butcher84} recipe, a constant 0.2 mag difference from the red sequence, should thus be updated with the actual knowledge that the universe was younger and more star forming at high ($z \sim 1$-2) redshift than it is today.
Basically, one should account for the secular evolution of stellar populations in the quenched fraction definition.
A possibility is to adopt a  (population synthesis) model for such an evolution \citep[e.g. following][who adopted a \citealt{bruzual03} $\tau=3.7$ Gyr]{andreon06}.
An even better solution is to determine the threshold with data, e.g. by locating the valley between the red sequence and the blue
cloud using the data as done at low redshift \citep[e.g.][]{baldry04} and at high redshift by \citet{andreon08a}.

% X-ray cluster selection
Several decades after \citet{butcher78a}, cluster galaxy evolution is still at the origin of a fruitful literature.
The wealth of available data now enables the building of enlarged cluster samples: to make the best use of those, it is of prime interest to understand to which extent the way clusters are selected may bias the observed galaxy evolution.
For instance, if clusters preferentially enter the sample at high redshift if they have a large fraction of blue galaxies, $f_{\rm{blue}}$ \citep[as in ][see \citealt{andreon99}]{butcher84}, or a dominant red population as in cluster searches based on them \citep[see][]{andreon06}.
It is clear that if galaxy populations are being studied, the cluster selection should as much as possible be independent of them (at a given cluster mass).
In this respect, X-ray cluster selection has the advantage that, at a given cluster mass, clusters are detected in X-ray independent of their $f_{\rm{blue}}$ (or quenching rate)\footnote{We emphasise that while $f_{\rm{blue}}$ may be correlated to the cluster X-ray luminosity, to invalidate the use of X-ray to select cluster requires that the correlation is at given cluster mass, while the observed correlation is largely driven by cluster mass.}.

% Cool-core
While the X-ray selection helps in making the sample selection independent of the quantity being measured, it is a truism to say that the X-ray detection of a cluster (of fixed mass) does depend on the X-ray properties of the cluster itself.
The presence of a cool-core/cooling flow \citep{fabian94} may favour the cluster detection because the source is X-ray brighter, or disfavour the cluster detection, at high redshift, when the source becomes hard to be distinguished from a point source.
It would be therefore interesting to know if cool-core clusters harbour different galaxy populations from non cool-core clusters (at a given cluster mass, galaxy stellar mass and normalised cluster-centric  distance), because of the potential selection effect suffered by X-ray cluster samples. 
Furthermore, since cool-core clusters are those more relaxed because the fragile cool-core is easily destroyed  by mergers, the comparison of such two cluster populations will inform us about the impact of recent cluster mergers on galaxy populations.

% Cluster mass
In addition, while it is certainly safe to compare local clusters to their likely ancestors, there would be two observational advantages to relax, at least in part, this requirement: on the one hand, a relaxed cluster mass selection would increase the number of clusters that can be studied; on the other hand, the stochastical nature of the mass growth in a hierarchical Universe (i.e. we know the mean mass evolution, but not the individual one) requires large cluster samples.
If galaxy populations strongly depend on cluster mass (more than via cluster-centric distance), then large cluster samples are needed to reduce the noise induced by the variety of the possible halo formation histories.
In particular, we emphasise that the control on cluster-centric distance (i.e. on $r/r_{200}$) might already fully capture the mass dependency, as cluster mass is defined to be $r^3_{200}$, apart from physical constants.
Moreover, besides selection effects, it is expected that different physical processes are efficient at different cluster masses \citep[e.g. see][]{moran07}.
It would therefore be interesting to know if the quenching rate depends on cluster mass, at a given (evolving) galaxy stellar mass and normalised cluster-centric distance.

% Metallicity
Eventually, cluster iron abundance also brings interesting constraints to cluster galaxy evolution, as metals produced by supernovae (Type Ia or Type II) within cluster galaxies and scattered into the intracluster medium \citep[more likely through galactic winds; e.g.][]{renzini93} cannot escape from the cluster gravitational well.
In this context, it is legitimate to study a possible correlation between the cluster iron abundance and the presence of star-forming cluster galaxies, which are expected to harbour massive stars exploding into supernovae.
 
 % This work
The dependence of quenching on hosting halo properties (mainly mass, richness, velocity dispersion, X-ray luminosity) has already been the subject of a fruitful literature \citep[e.g.][to cite a few]{de-propris04,goto05,wake05,poggianti06,aguerri07,popesso07,haines09,balogh10}.
However, because of the difficulty in gathering suitable samples with homogeneous data, the majority of those analyses are done only for one bin of galaxy stellar mass and environment.
Only very few analyses \citep[e.g.][]{weinmann06,peng12} have been led as a function of galaxy stellar mass and environment at the same time.
\citet{weinmann06} found a dependence on halo mass when studying log($M_{200}/M_\odot) > 11.5$ systems, while \citep{peng12} do not observe any dependence on halo mass for log($M_{200}/M_\odot) > 13$ systems.

The present work thus aims to investigate how the cluster properties are related to cluster galaxy evolution, with leading the analysis as a function of galaxy stellar mass and cluster-centric distance.
We use an X-ray selected cluster sample (20 clusters with $0.02 \le z \le 0.1$ and log($M_{200}/M_\odot) \ge 13.9$) to probe if there is any dependence of the quenching rate in clusters with the hosting cluster properties (mass, richness, iron abundance, and central cooling time), and use the fraction of blue galaxies $f_{\rm{blue}}$ as a proxy for quenching.

% Paper outline
The plan of this paper is as follows.
In Sect. \ref{sec:data} we define the cluster sample on which this study relies, the data and their analysis, including our estimation of $f_{\rm{blue}}$.
Our results are presented in Sect. \ref{sec:results} and Sect. \ref{sec:logMrfit}, and summarised in Sect. \ref{sec:conclusion}.

% Paper conventions
We adopt $H_0 = 70$ km s$^{-1}$ Mpc$^{-1}$,  $\Omega_m = 0.30$, and $\Omega_\Lambda = 0.70$ throughout.
Galaxy stellar masses are computed with a \citet{chabrier03} initial mass function and are defined by the mass of the gas that will eventually be turned into stars, i.e. corresponding to the integral of the star formation rate over $0 \le t < \infty$, \citep[as in][]{andreon08a,raichoor12,raichoor12b}.

% Bayesian approach
All our results ($f_{\rm blue}$ values and the fitting with Eq.(\ref{eq:model})) are obtained through a Bayesian approach using Markov Chain Monte Carlo (MCMC) simulations.
For the stochastic computation and for building the statistical model, we use Just Another Gibb Sampler \citep[\textsc{jags}\footnote{\url{http://mcmc-jags.sourceforge.net/}},][]{plummer10}, which is a program for analysis of Bayesian hierarchical models using MCMC simulation.
Readers not familiar with Bayesian methods may think of our approach as close to a maximum likelihood fit of unbinned data.
For each quantity, the quoted 68\% confidence interval represents the shortest interval, including 68\% of the computed posterior values for this quantity.

%@@@@@@@@@@@@@@@@@@@@@@@@@@@@@@@@@@@@@@@@@@@
% CLUSTER SAMPLE AND FBLUE FRACTION
%@@@@@@@@@@@@@@@@@@@@@@@@@@@@@@@@@@@@@@@@@@@

\section{Cluster sample and $f_{\rm{blue}}$ estimation \label{sec:data}}

\subsection{Cluster sample and cluster properties}

In this section we describe our cluster sample, which is an X-ray selected low-redshift sample.
Our aim being to study galaxy properties through their colour, we require a cluster sample selected on a criterion independent of galaxy colour: X-ray selection ensures that our cluster sample selection is unbiased in $f_{\rm{blue}}$, because the probability of inclusion of a cluster in the sample is independent of $f_{\rm{blue}}$ at a given cluster mass.
We start from the X-ray flux-limited HIghest X-ray FLUx Galaxy Cluster Sample \citep[HIFLUGCS,][]{reiprich02}.
We apply a spatial selection to ensure an optical photometric coverage by the Sloan Digital Sky Survey \citep[SDSS;][]{york00} and a redshift ($0.02 \le z \le 0.1$) selection to avoid shredding and ensure a decent signal-to-noise ratio in the SDSS photometry.
We thus obtain 20 low-redshift clusters (MKW4, A1367, A1656, MKW8, A2199, A2634, A2052, A2147, A2063, A2657, A2589, MKW3S, A1795, A2065, ZwCl1215, A2029, A2255, A1650, A2142, A2244).
We note that the seven less massive of those low-redshift clusters are included in the \citet{raichoor12b} sample.

From \textit{Chandra} X-ray observations, \citet{hudson10} derived cluster temperatures $T_X$, metallicities $Z$, and central cooling times $CCT$.
We derive $r_{200}$ for all the clusters from $T_X$ \citep{finoguenov01,willis05,raichoor12b},  which is of paramount importance, because it is known that $f_{\rm{blue}}$ depends on cluster-centric distance.
By definition, the cluster mass within the overdensity 200 (virial) is:
\begin{equation}
M_{200} [\textnormal{M}_\odot] = \frac{4}{3} \pi \times (r_{200}[\textnormal{Mpc}])^3 \times 200 \rho_c(z)[\textnormal{M}_\odot \, \textnormal{Mpc}^{-3}],
\label{eq:lgM200}
\end{equation}
where $\rho_c(z) = 3 H_0^2 \times (\Omega_m(1+z)^3+\Omega_\Lambda)/(8 \pi G)$ is the critical density at the redshift $z$.
We stress that using another definition for the cluster mass \citep[e.g.][]{pacaud07,hudson10} has little impact because our sample is at low redshift ($z  \le 0.1$), and does not affect our conclusions.

An additional measurement of the halo properties is the cluster richness, derived as a byproduct of our computation of $f_{\rm{blue}}$, for each bin of cluster-centric distance $r/r_{200}$ and galaxy stellar mass $M$ (Sect. \ref{sec:fblue}; see Sect. 4.2 of \citealt{raichoor12} for more details).
Hence we can estimate the cluster richness $N_{200}$ by simply adding the number of blue and red cluster galaxies (more massive than $10^{10.73} M_\odot$ and within $r_{200}$) .

\begin{figure}
\includegraphics[width=\linewidth]{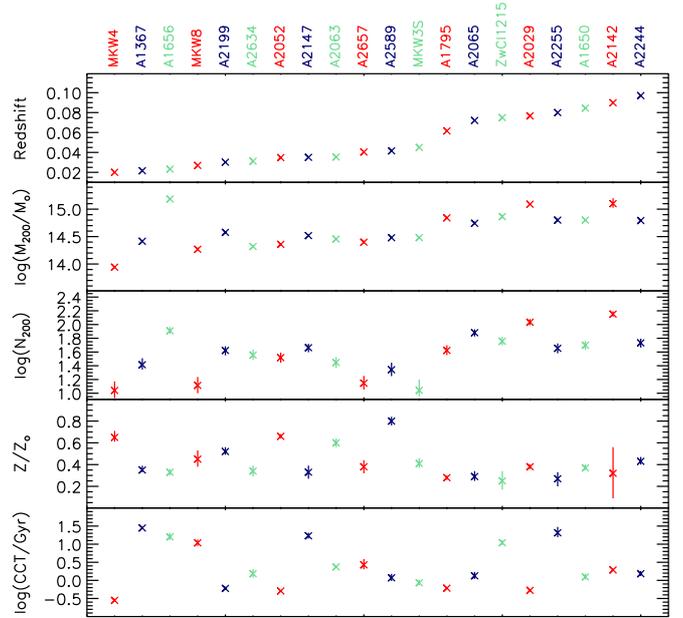}
\caption{
Cluster sample properties (from top to bottom): redshift $z$, mass log($M_{200}/M_\odot$), richness log($N_{200}$), iron abundance $Z/Z_\odot$ and central cooling time log($CCT$/Gyr) (\citealt{hudson10}; the richness comes from the present work).
\label{fig:cluster_properties}
}
\end{figure}

Figure \ref{fig:cluster_properties} displays the properties of our cluster sample.
% mass
Our cluster sample has masses within $13.9 \le \textnormal{log} (M_{200}/M_\odot) \le 15.2$, thus including massive clusters.
As our cluster sample is X-ray flux-selected, there is a correlation between redshift and mass; however, the redshift range spans only $\sim$1 Gyr ($0.02 \le z \le 0.1$).
% richness
Richnesses are within $1.0 \le \textnormal{log}(N_{200}) \le 2.2$, and the well-known cluster mass-richness correlation \citep[e.g.][]{koester07,andreon10b} is clearly visible.
% metallicity 
Regarding iron abundance, our sample spans a typical range of $0.25 \le Z/Z_\odot \le 0.80$.
% CCT
Finally, central cooling times are within $-0.6 \le \textnormal{log} (CCT/Gyr) \le 1.4$, thus spanning the typical range of cooling times: our sample is composed of approximately an equal number of strong cool-core ($\textnormal{log} (CCT/Gyr) < 0$), weak cool-core ($0 < \textnormal{log} (CCT/Gyr) < 0.9$), and non-cool-core ($0.9 < \textnormal{log} (CCT/Gyr)$) clusters, using \citet{hudson10} nomenclature.
This cluster sample spans the typical cluster values for iron abundance and central cooling time.
It is thus representative of massive clusters and allows us to probe the dependence of $f_{\rm{blue}}$ on those four quantities.

% FIGURE: cluster properties

%@@@@@@@@@@@@@@@@@@@@@@@@@@@@@@@@@@@@@@@@@@@
% FBLUE COMPUTATION
%@@@@@@@@@@@@@@@@@@@@@@@@@@@@@@@@@@@@@@@@@@@

\subsection{The $f_{\rm{blue}}$ estimation \label{sec:fblue}}

We evaluate the quenching rate in clusters as a function of galaxy stellar mass $M$ and cluster-centric distance $r/r_{200}$, through the fraction of blue galaxies, $f_{\rm{blue}}$, which is estimated with the procedure described in \citet{raichoor12,raichoor12b}, using the SDSS DR9 data \citep{ahn12}.
We used the \texttt{ModelMag} for galaxy colors and \texttt{cModelMag} as an approximate total magnitude for galaxies.
Our background contamination estimation ensures $0 \le f_{\rm{blue}} \le 1$, and our $f_{\rm{blue}}$ estimation accounts for stellar evolution of galaxies with time, which is minimal here because of the considered redshift window ($0.02 \le z \le 0.1$).

We summarise the main steps of the procedure for each cluster here:
\begin{itemize}
\item we worked in a $(u-r)$ vs $r$ colour-magnitude diagram, thus probing the 4000 \AA~ break;
\item we considered three radial bins (annuli) at distinct cluster-centric radii, defined by $r/r_{200} \le 0.5$ , $0.5 < r/r_{200} \le 1$, and $1 < r/r_{200} \le 2$;
\item we considered four galaxy stellar mass bins using loci in colour-magnitude diagrams
($\log (M/M_{\sun})$ in [9.92, 10.33], [10.33, 10.73], [10.73, 11.13], and $>11.13$).
Those loci are defined using the 2007 version of \citet{bruzual03} model (CB07; solar metallicity, formation redshift of $z_{form}=5$), with an exponentially declining star-forming $\tau$ model with $0 \le \textnormal{SFH} \; \tau \; \textnormal{(Gyr)} \le 10$ \citep[e.g. see Figure 8 of][]{raichoor12b}.
We hereafter refer to those galaxy stellar mass bins using the mean value of each bin (weighted by a \citealt{schechter76} function), i.e. $\langle \log (M/M_{\sun})  \rangle \sim 10.14$, 10.54, 10.94, and 11.47.
We required that the lowest galaxy stellar mass cut is brighter than a signal-to-noise ratio of 5 in the considered colour-magnitude diagram;
\item we removed the fore/background galaxies in the cluster line of sight in a two-step procedure, first using a minimal individual background subtraction.
We use the full photometric redshift probability distribution function $p(z)$ estimated with \textsc{Eazy} \citep{brammer08} (with a $r$-band prior, see Appendix B of \citealt{raichoor12b}) to remove galaxies that are definitely in front of or behind the cluster, and then we use an independent surrounding control sample ($\sim$90 deg$^2$) to statistically account for the residual background;
\item we defined a galaxy as blue if it is bluer than a CB07 model with $\tau = 3.7$ Gyr, \citep[as in][]{andreon04,raichoor12,raichoor12b}; such a definition is consistent with the original definition of \citet{butcher84} at $z=0$;
\item for each galaxy stellar mass bin and radial bin we computed the blue fraction $f_{\rm{blue}}$, which accounts for residual background galaxies using our control samples, following the Bayesian methods introduced in \citet[][see their Appendix C]{andreon06}.
In short, we build with \textsc{jags} MCMC simulations to compute the probability of each value of the blue fraction, given the data, using the Bayes theorem of statistics; we account for the fact that the observed number of blue (total) galaxies are drawn from an underlying binomial (Poissonian) distribution, and proceed similarly to model background galaxy counts.
\end{itemize}

%@@@@@@@@@@@@@@@@@@@@@@@@@@@@@@@@@@@@@@@@@@@
% RESULTS
%@@@@@@@@@@@@@@@@@@@@@@@@@@@@@@@@@@@@@@@@@@@

\section{Results: dependence of $f_{\rm{blue}}$ with cluster properties \label{sec:results}}

We have 204 individual measurements of $f_{\rm{blue}}$ in 20 low-redshift massive clusters, covering four bins of galaxy stellar mass ($10.14 \le \langle \textnormal{log}(M/M_\odot) \rangle \le 11.47$) and three bins of cluster-centric distance ($0 \le r/r_{200} \le 2$), which are displayed in Figure \ref{fig:fblue_lgM200_indiv}.
In this section, we are looking for the effect on top of the well-known dependence of the quenched fraction on galaxy stellar mass and environment.
We test the dependence of $f_{\rm{blue}}$ on cluster-related quantities denoted below by $Q$: cluster mass $\textnormal{log}(M_{200}/M_\odot)$, cluster richness $\textnormal{log}(N_{200})$, cluster iron abundance $Z/Z_\odot$, and cluster central cooling time $\textnormal{log}(CCT/\textnormal{Gyr})$.
We recall that our measurements of $f_{\rm{blue}}$ already accounts for the cluster size (by normalising the cluster-centric distance to $r_{200}$), hence at first order for quantities as cluster mass or richness.

\subsection{The $f_{\rm{blue}}$ model fitting \label{sec:results_fbluefit}}

It is well established that the quenching rate depends on the galaxy stellar mass $M$ \textit{and} on the cluster-centric distance $r/r_{200}$ \citep[e.g.][]{peng10,raichoor12b}.
Therefore, any modeling of the dependence on a cluster quantity should, at the very least, first control for these two covariates, i.e. also explicitly model these two quantities.
In other terms, for each cluster quantity $Q$, we need to fit at once the dependency on $Q$, on galaxy stellar mass $M$, and on cluster-centric radius $r/r_{200}$.

Following the approach of \citet{raichoor12b} for each quantity $Q$ in $\{$log$(M_{200}/M_\odot)$, log$(N_{200})$, $Z/Z_\odot$, log$(CCT/\textnormal{Gyr})\}$, we analyse the simultaneous dependence of $f_{\rm{blue}}$ on $\{r/r_{200},M, Q\}$, \textit{by fitting the 204 individual measurements of $f_{\rm{blue}}$} with a function having the following form:
\begin{eqnarray}
{f_{\rm{blue}}} \left(r/r_{200},M,Q \right) &=& \texttt{ilogit} \, \Bigl[ A_{0,Q} + \nonumber\\
&&	\alpha_Q \cdot \log \left(r/(0.25 \cdot r_{200}) \right) + \nonumber \\
&&	\beta_Q \cdot (\log(M/M_{\sun}) - 11) + \nonumber \\
&&	\gamma_Q \cdot (Q-Q_0) \Bigr],
\label{eq:model}
\end{eqnarray}
where $\texttt{ilogit}(x) = (1+\exp(-x))^{-1}$ ensures that $0 \le f_{\rm{blue}} \le 1$.
We adopt uniform priors for the parameters $A_{0,Q}$, $\alpha_Q$, $\beta_Q$, and $\gamma_Q$.
The only motivation behind the chosen parametrisation is the adoption of an additive model that fits our data and ensures that $0 \le f_{\rm{blue}} \le 1$.
For clarity, we do not include the redshift $z$ in the fit, as our cluster sample spans a relatively small redshift interval ($0.02 \le z \le 0.1$), and including it does not change our results.

No matter which quantity $Q$ we are fitting, three parameters have the same value in all fittings:
$A_{0,Q} = -4.2^{+0.3}_{-0.3}$, 
$\alpha_Q=1.3^{+0.4}_{-0.4}$, 
and $\beta_Q=-2.6^{+0.4}_{-0.4}$.
The fitted coefficients $\gamma_Q$ (along with the fixed values $Q_0$) are displayed in Table \ref{tab:fit_results} and we present in Figure \ref{fig:gamma_post} their computed posterior.

% TABLE: CLUSTER SAMPLE
\begin{table*}
\centering
\caption{Fitted coefficients of Eq.(\ref{eq:model}) for our cluster sample. \label{tab:fit_results}}
\begin{tabular}{l l c c c c}
\hline
\hline
$Q$								&	Data range in $Q$	&	$Q_0$	&	$\gamma_{Q}$		&	mean($|\Delta f_{\rm{blue}}|$)\\
\hline
$\textnormal{log}(M_{200}/M_\odot)$	&	[13.9, 15.2]		&	14.5		&	$0.5^{+0.4}_{-0.4}$	&	0.02\\
$\textnormal{log}(N_{200})$ 			&	[1.0, 2.2]			&	1.8		&	$0.7^{+0.6}_{-0.4}$	&	0.03\\
$Z/Z_\odot$						&	[0.25, 0.80]		&	0.4		&	$0.4^{+1.0}_{-1.0}$	&	0.01\\
$\textnormal{log}(CCT/\textnormal{Gyr})$	&	[-0.5, 1.5]			&	0.5		&	$-0.1^{+0.2}_{-0.2}$	&	0.00\\
\hline
\end{tabular}
\tablefoot{
The coefficients $\gamma_0$ and $Q_0$ refer to Eq. (\ref{eq:model}).
mean($|\Delta f_{\rm{blue}}|$) represent the expected change in the probability that a galaxy is quenched implied by the mean relation ignoring the $Q$-dependence (see text for more details).}
\end{table*}

To visualise how the model fits the data, we display in Figures \ref{fig:fblue_lgM200}--\ref{fig:fblue_lgCCT} the variations of $f_{\rm{blue}}$ when we fix $r/r_{200}=0.75$ or log$(M/M_\odot)=10.54$.
For each cluster quantity $Q$, we present the stacked data\footnote{We note that the expected uncertainty in log($M_{200}/M_\odot$) due to the uncertainty in $T_X$ and on the  $M_{200}-T_X$ relation is about 0.2 dex \citep[e.g.][]{mahdavi13} and is thus smaller than our bin width in log($M_{200}/M_\odot$) ($>$0.3 dex).
In addition, the uncertainty in the other quantities ($\sigma_{\textnormal{log}(N_{200})} \sim 0.1$ dex, $\sigma_{Z/Z_\odot} \sim 0.05$, and $\sigma_{\textnormal{log}(CCT/Gyr)} \sim 0.1$ dex) are significantly smaller than our bin width ($\sim 0.5$ for log$(N_{200})$ dex, $\sim 0.3$ for $Z/Z_\odot$, and $\sim 1$ dex for log$(CCT/Gyr)$).} and the fitted model in two complementary manners: as a surface to show the overall behaviour of $f_{\rm{blue}}$ and as slices for fixed values of $\{$log$(M/M_\odot)$, $r/r_{200}\}$, where we display the model uncertainty.
The individual data are stacked only for display purpose, so that the trends can stand out; we recall that the fit was performed on the individual 204 values of $f_{\rm{blue}}$.\\

We remark that our model satisfactorily fits the data, which means that its simple structure is sufficiently flexible.
As found in previous works \citep{peng10,raichoor12b}, there is no need for a term crossing the galaxy stellar mass $M$ and the cluster-centric distance $r/r_{200}$, hence the mass quenching and the environmental quenching are separable (see also Sect. \ref{sec:logMrfit}).

For each quantity $Q$, we observe that the fitted model has no significant dependence on $Q$, for the range in $Q$ probed by our data (Column 2 of Table \ref{tab:fit_results}).
Indeed, a null slope $\gamma$ is almost always included within the 68\% confidence interval.
To quantify the implication of the lack of a significant dependence, we compute the change in the probability that a galaxy is quenched implied by the mean relation ignoring the $Q$-dependence.
In practice, we compute  $|\Delta f_{\rm{blue}}|$, the predicted (best fit) $f_{\rm blue}$ change on a range given by the 68\% interval of $Q$ per each galaxy stellar mass and radial bin, and take their average.
We use the mean of $|\Delta f_{\rm{blue}}|$ because the maximum values are always reached where the data and model uncertainties are the largest).
We report the mean value of $|\Delta f_{\rm{blue}}|$ in the last column of Table \ref{tab:fit_results}.
Hence, these halo properties change, on average, by less than $<3\%$ the probability that a galaxy is quenched, once the mass--size ($M_{200}$ -- $r_{200}$) scaling relation is accounted for through cluster-centric distance normalisation. 
We now detail the analysis for each quantity $Q$.

% Mass and richness
\subsection{Cluster mass log($M_{200}/M_\odot$) and richness log($N_{200}$)}
Some cluster-related quenching processes, such as ram-pressure stripping, are linked to the cluster mass, which thus may be correlated to $f_{\rm{blue}}$.
Our computation of $f_{\rm{blue}}$ accounts at first order for the cluster mass by normalising the cluster-centric distance to $r_{200}$.
In addition, such processes have an efficiency varying with the cluster-centric distance \citep[e.g.][]{treu03} and are in competition with quenching processes internal to the galaxy, hence depending on galaxy stellar mass.

\begin{figure*}
\begin{tabular}{@{\extracolsep{5pt}} cc}
	\includegraphics[width=0.5\linewidth]{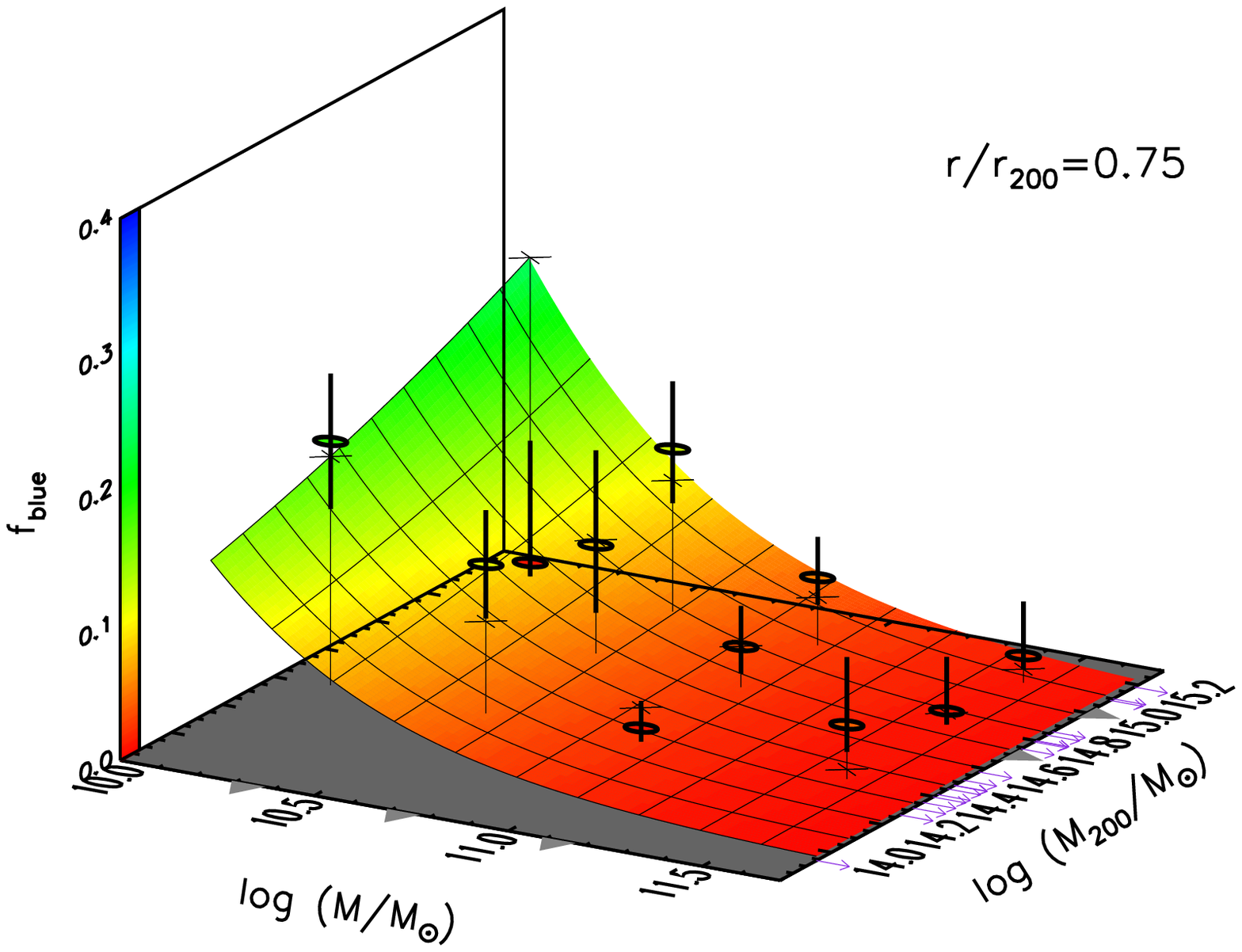} &
	\includegraphics[width=0.5\linewidth]{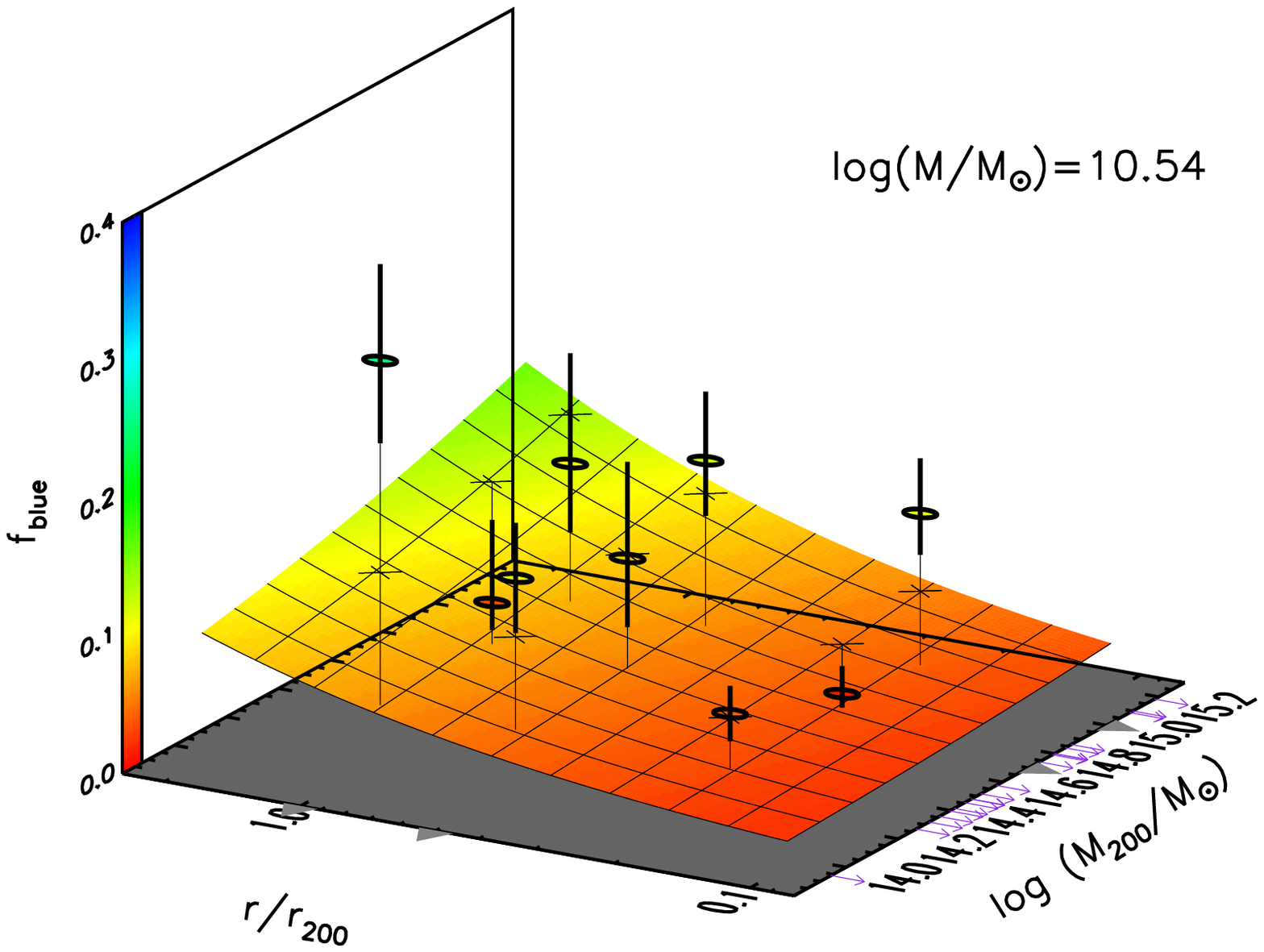} \\
	&\\
	\includegraphics[width=0.5\linewidth]{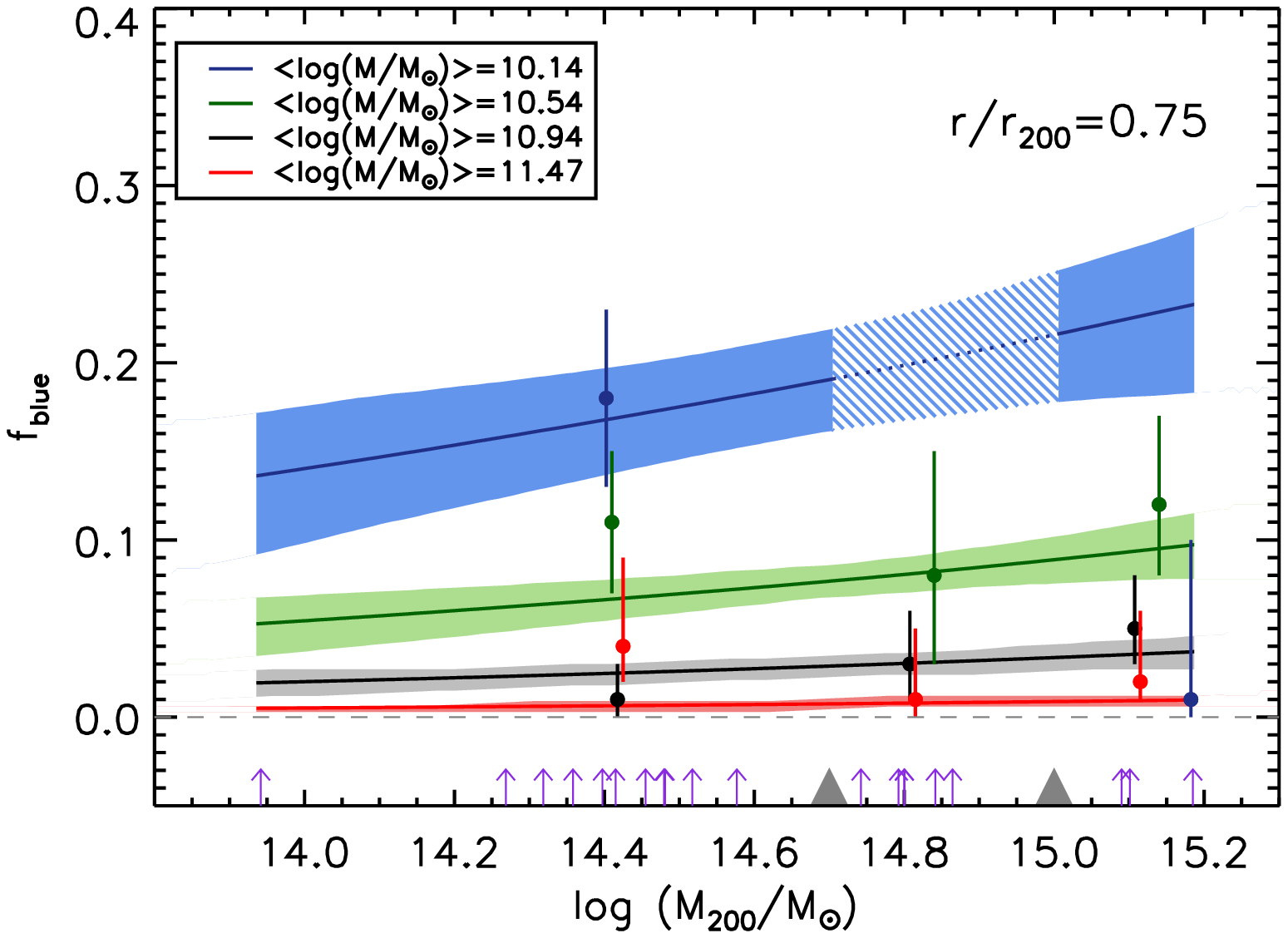} &
	\includegraphics[width=0.5\linewidth]{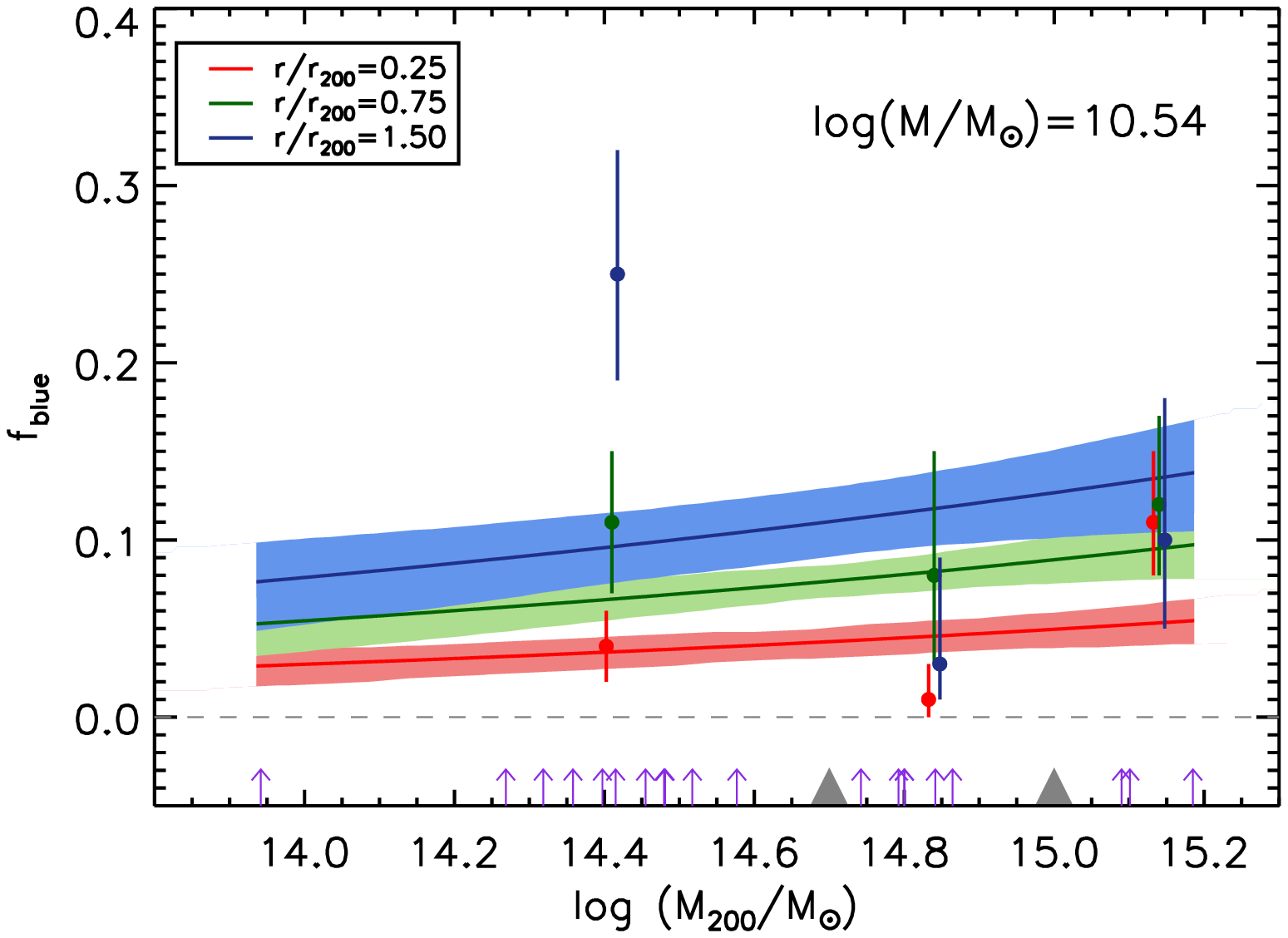}\\
\end{tabular}
\caption{
Two-dimensional slice/view of the six-dimensional parameter space describing the dependence of the quenching rate on galaxy and cluster properties.
The figures consider the following quantities: cluster mass log$(M_{200}/M_\odot)$, galaxy stellar mass $M$, and cluster-centric distance $r/r_{200}$.
The left (right) column consider $r/r_{200}=0.75$ (log($M/M_\odot)=10.54$).
The bin boundaries are indicated by grey filled triangles on the axis.
Error bars represents the shortest interval including 68\% of the posterior values.
Individual cluster sample log$(M_{200}/M_\odot)$ values are indicated by purple arrows.
\textit{Upper panels}: Thick circles with error bars represent the stacked data and the error.
The colour-coded surface represents the fitted model (mean value of the posterior distribution at each [x,y] locus).
The model value of $f_{\rm{blue}}$ at the [x,y] locus of each stacked data is symbolised by a cross.
\textit{Lower panels}: Data points with error bars represent the stacked data (slightly shifted along the x-axis for clarity).
Solid lines and shaded areas represent the posterior mean and error of the model of Eq. (\ref{eq:model}).
Prediction/extrapolation of this model for bins where we do not have data are plotted as dotted lines and hatched areas.
The fit is performed on the 204 individual $f_{\rm{blue}}$ measurements.
\label{fig:fblue_lgM200}
}
\end{figure*}

We present in Figure \ref{fig:fblue_lgM200} how $f_{\rm{blue}}$ depends on galaxy stellar mass log($M/M_\odot$), cluster-centric distance $r/r_{200}$, and cluster mass log($M_{200}/M_\odot$).
We find a slope $\gamma_{M_{200}}=0.5^{+0.4}_{-0.4}$, i.e. indicating no significant dependence.
The average implied variation of $f_{\rm{blue}}$ over our typical range of $M_{200}$ is 0.02, i.e. negligible.
Hence we can consider that $f_{\rm{blue}}$ does not significantly depend on cluster mass for our probed cluster mass range (log$(M_{200}/M_\odot) \gtrsim 14$).
To illustrate the fitting with the individual values, we display in Figure \ref{fig:fblue_lgM200_indiv} $f_{\rm{blue}}$ for the whole dataset as a function of cluster-centric distance, galaxy stellar mass, and cluster mass.

\begin{figure*}
\begin{tabular}{@{\extracolsep{5pt}} cc}
	\includegraphics[width=0.5\linewidth]{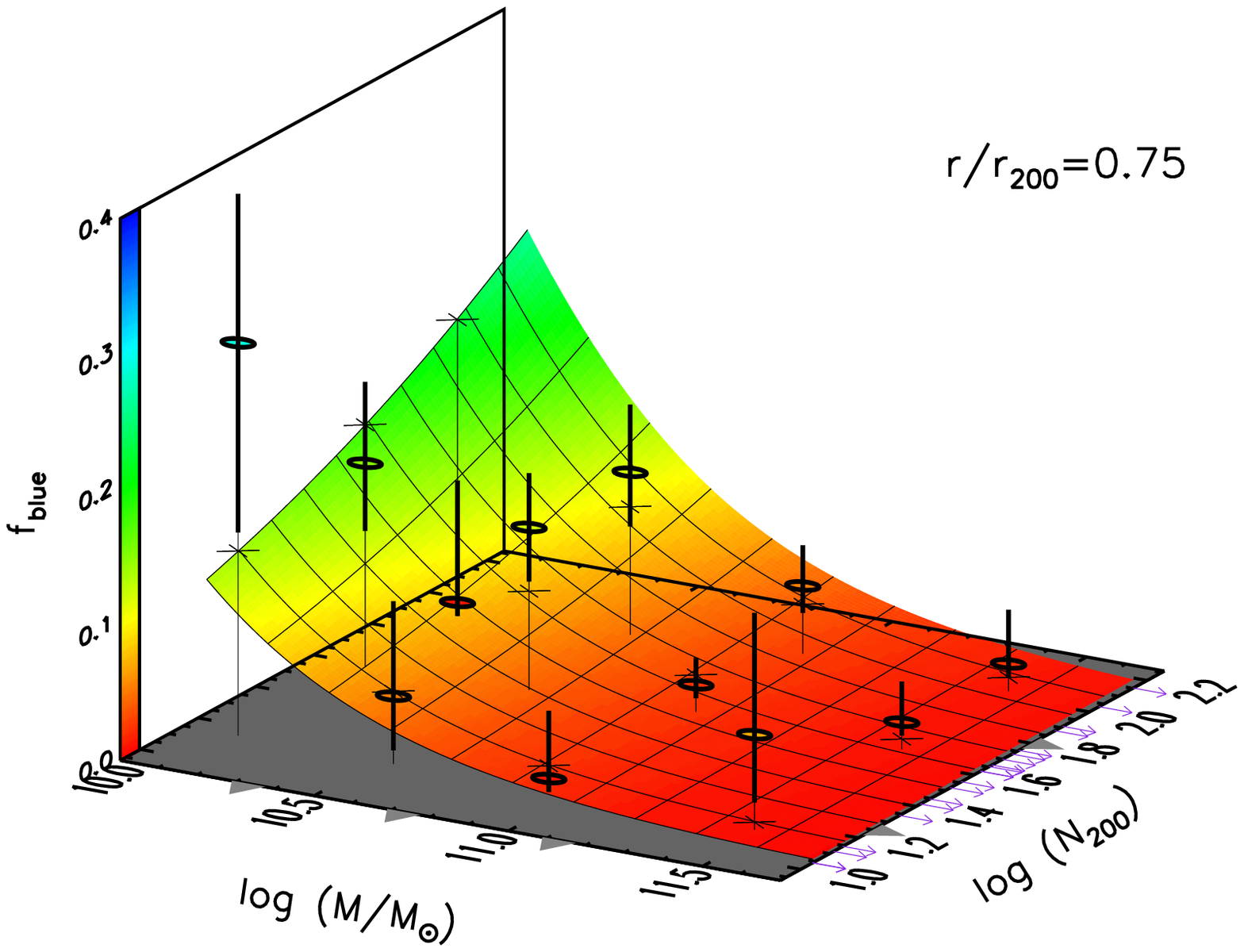} &
	\includegraphics[width=0.5\linewidth]{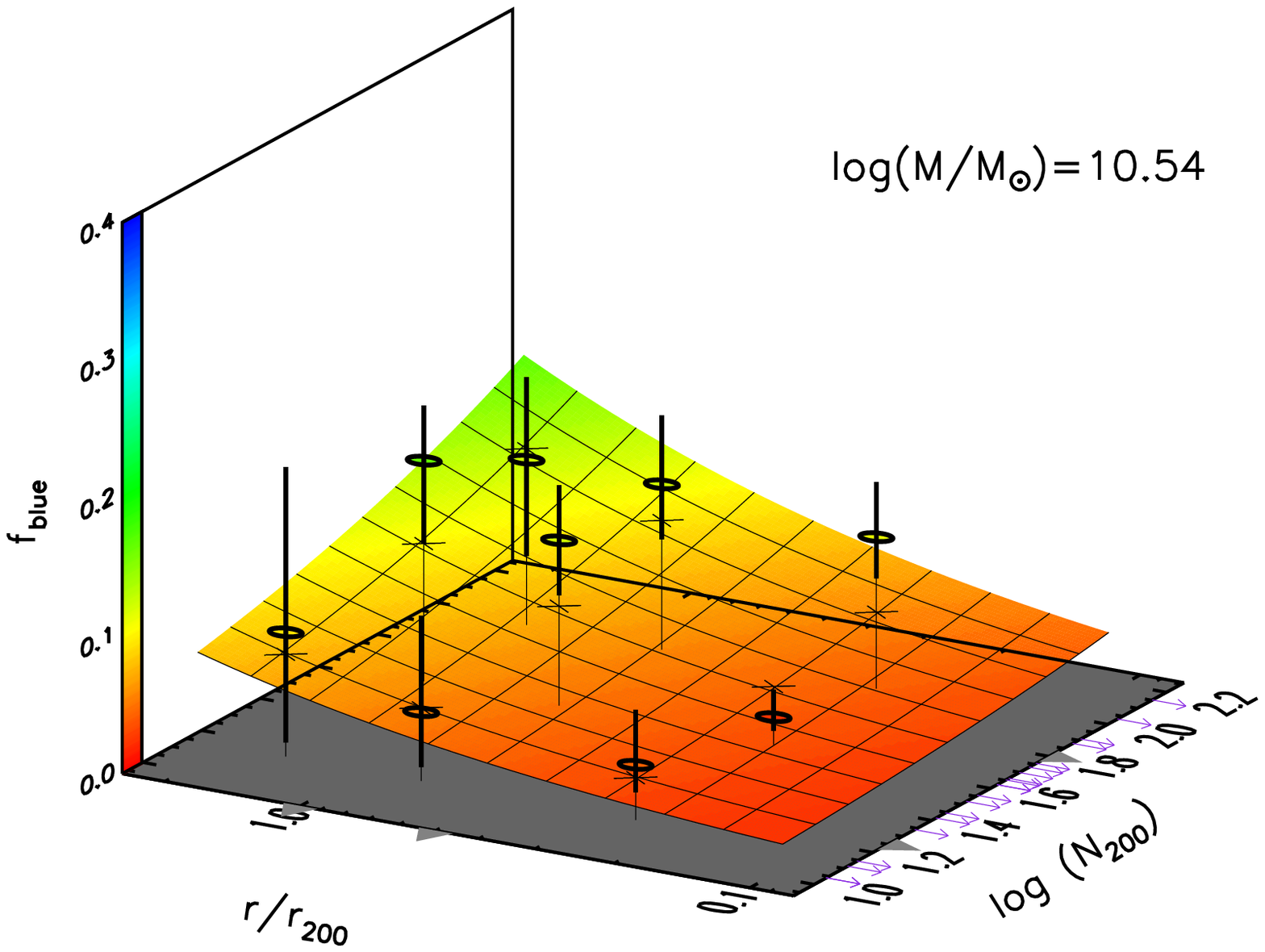} \\
	&\\
	\includegraphics[width=0.5\linewidth]{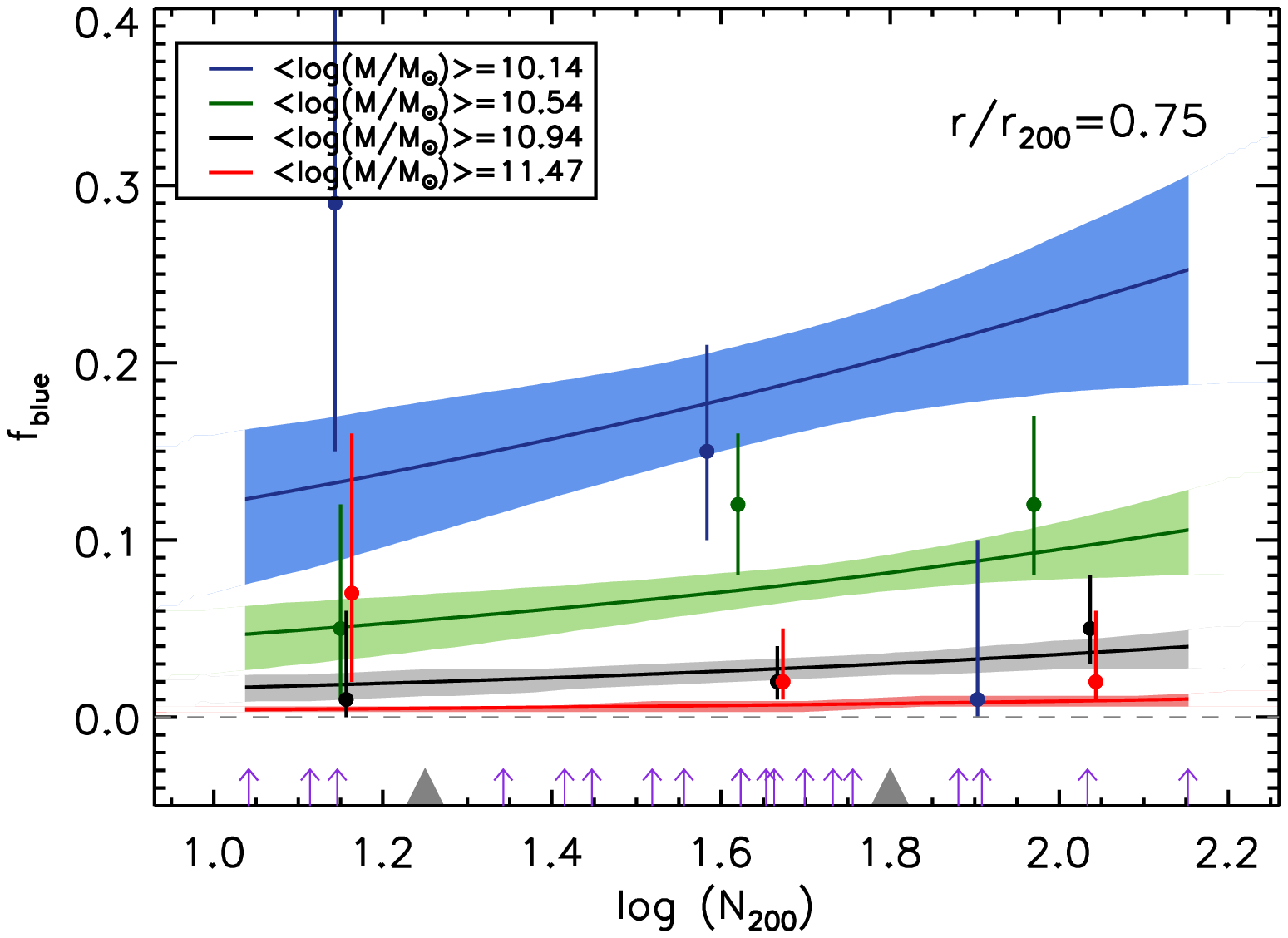} &
	\includegraphics[width=0.5\linewidth]{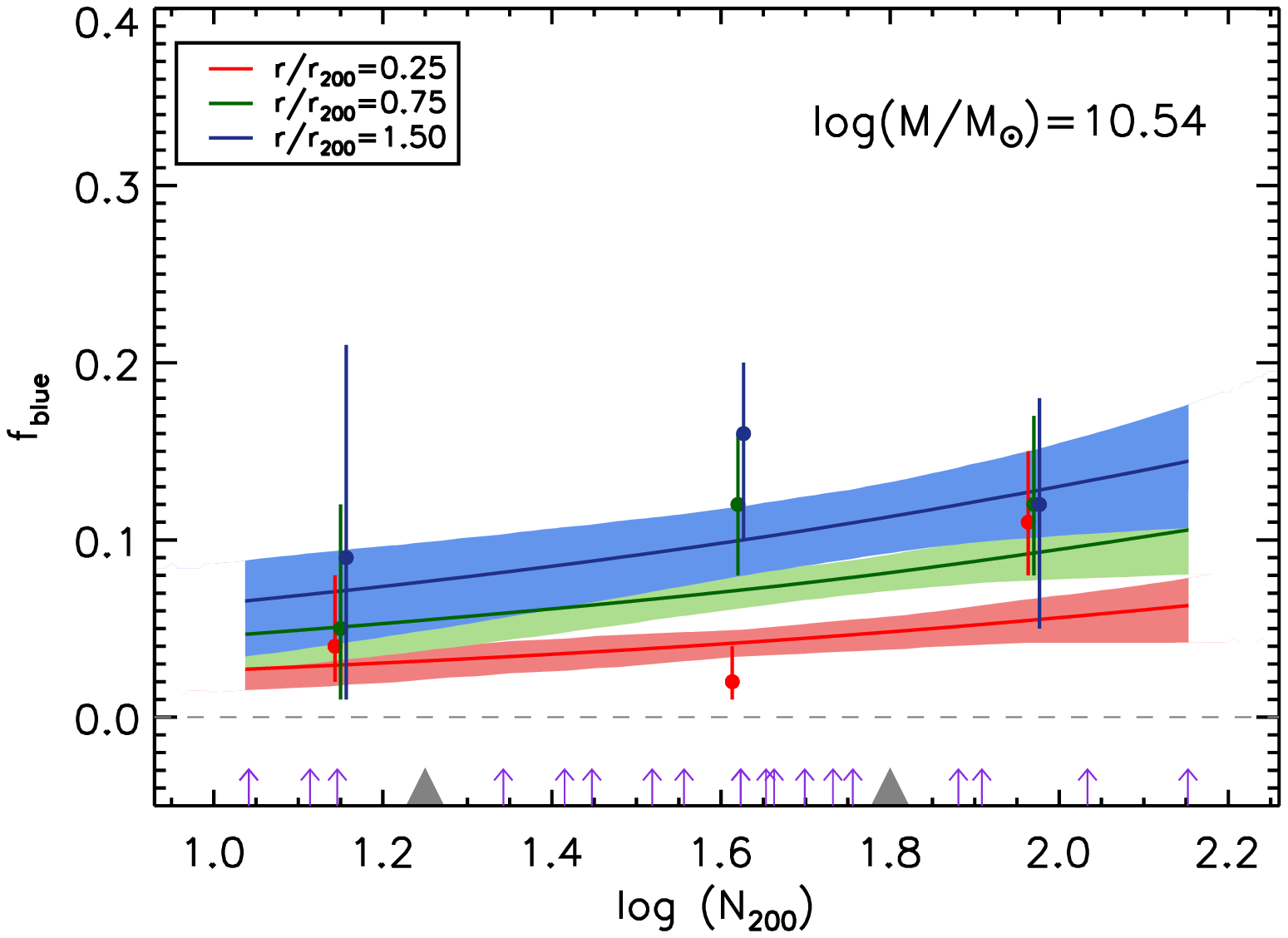}\\
\end{tabular}
\caption{
Two-dimensional slice/view of the six-dimensional parameter space describing the dependence of the quenching rate on galaxy and cluster properties.
The figures consider the following quantities: cluster richness log$(N_{200})$, galaxy stellar mass $M$, and cluster-centric distance $r/r_{200}$.
The fit is performed on the 204 individual $f_{\rm{blue}}$ measurements.
Symbols are similar to those in Figure \ref{fig:fblue_lgM200}.
\label{fig:fblue_lgN200}
}
\end{figure*}

Cluster richness is closely linked to cluster mass: we find a slope $\gamma_{N_{200}}=0.7^{+0.6}_{-0.4}$ (see Figure \ref{fig:fblue_lgN200}).
As expected, this slope is very close to $\gamma_{M_{200}}$, the slope of $f_{\rm{blue}}$ with respect to cluster mass, as those two properties are correlated with a slope close to unity.
Hence, we do not find any significant dependence of $f_{\rm{blue}}$ on cluster richness either.
We remark that \citet{peng12}, who led an analysis qualitatively similar to ours, reaches a similar conclusion.

The lack of an halo-mass dependency agrees with previous works \citep{de-propris04,goto05,aguerri07,haines09,balogh10,peng12}.
However, except for \citet{peng12}, these works do not split galaxies in galaxy stellar mass bins and cluster-centric distances.
This approach runs the risk of missing  halo-mass dependency because of the presence of opposite trends for different galaxy stellar masses or cluster-centric distances, or of a galaxy stellar mass term compensating a cluster-centric distance term.
Our work minimises this risk by explicitly accounting for galaxy stellar mass and cluster-centric distance.
Furthermore, our work places an unrivalled stringent upper limit on the effect of halo on the probability that a galaxy is quenched (once the mass-size $M_{200}$ -- $r_{200}$ scaling relation is accounted for through cluster-centric distance normalisation): this effect may change the probability by 3\% in most cases (Table \ref{tab:fit_results}).

\citet{weinmann06} find halo-mass dependency, however this dependency is at cluster masses lower than studied in our work.
It would be interesting to extend our work to the range of masses explored by \citet{weinmann06} because they used noisy masses derived from an abundance matching technique, whose noisiness also may hide an existing halo-mass dependency or induce a fake one.
Furthermore, their sample, unlike our sample, is selected dependent on the quantity being studied (galaxy properties), making it prone to selection effects.

% FIGURE: FBLUE = f(CLUSTER MASS)

% FIGURE: FBLUE = f(CLUSTER RICHNESS)

% Metallicity
\subsection{Cluster iron abundance $Z_0$}
Most of the metals produced within cluster galaxies are scattered within the intracluster medium, where they are retained because of the gravitational potential.
The cluster iron abundance thus keeps an imprint of the star formation history of the cluster galaxies.
The link between the metals produced within cluster galaxies and the cluster iron abundance is complex and has been debated for more than two decades (e.g. see \citealt{renzini08} and \citealt{renzini14} for an overview).
For instance, a cluster iron abundance increasing with the presence of star-forming cluster galaxies would imply a significant role of Type II supernovae in the current enrichment of the intracluster medium.

\begin{figure*}
\begin{tabular}{@{\extracolsep{5pt}} cc}
	\includegraphics[width=0.5\linewidth]{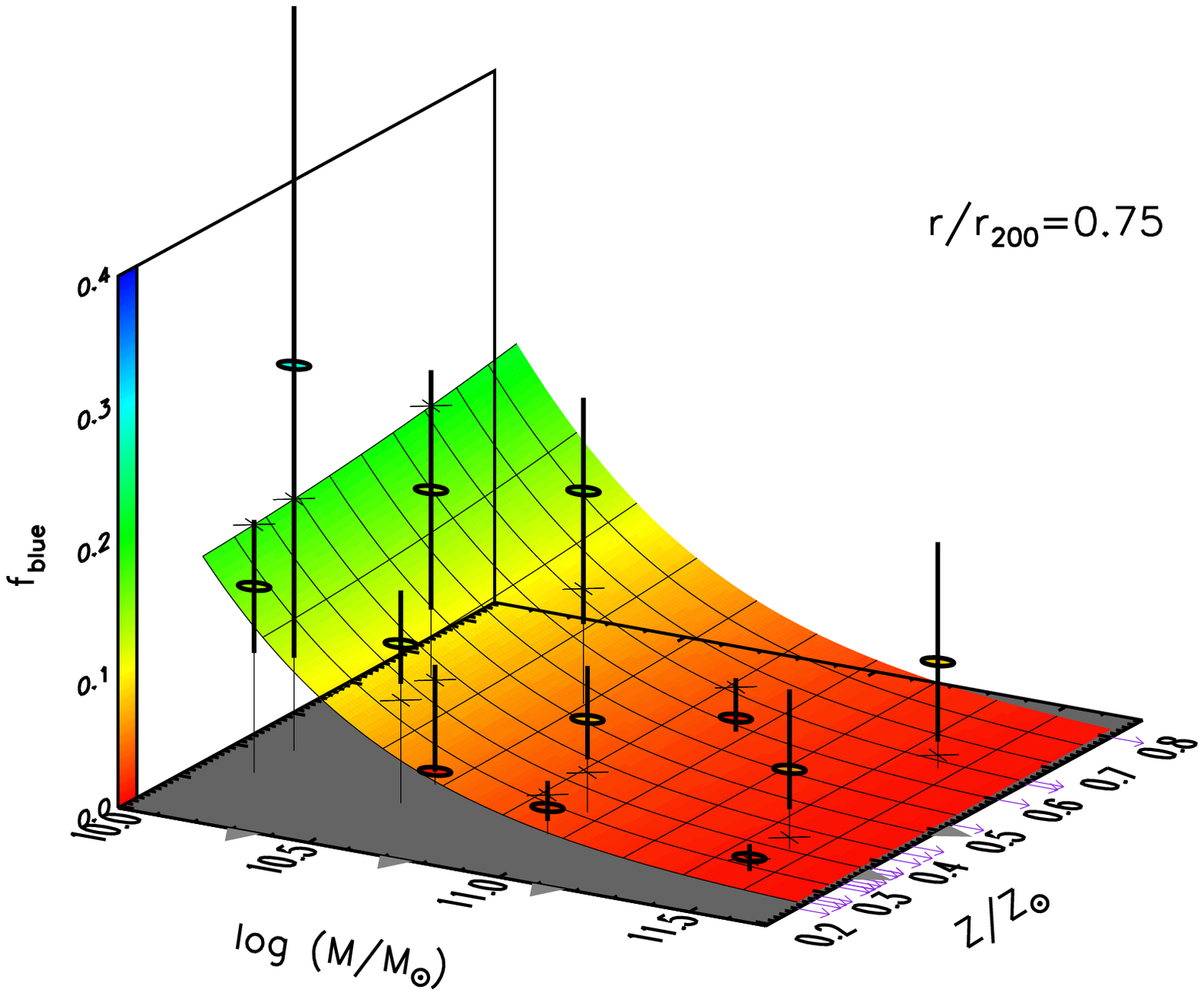} &
	\includegraphics[width=0.5\linewidth]{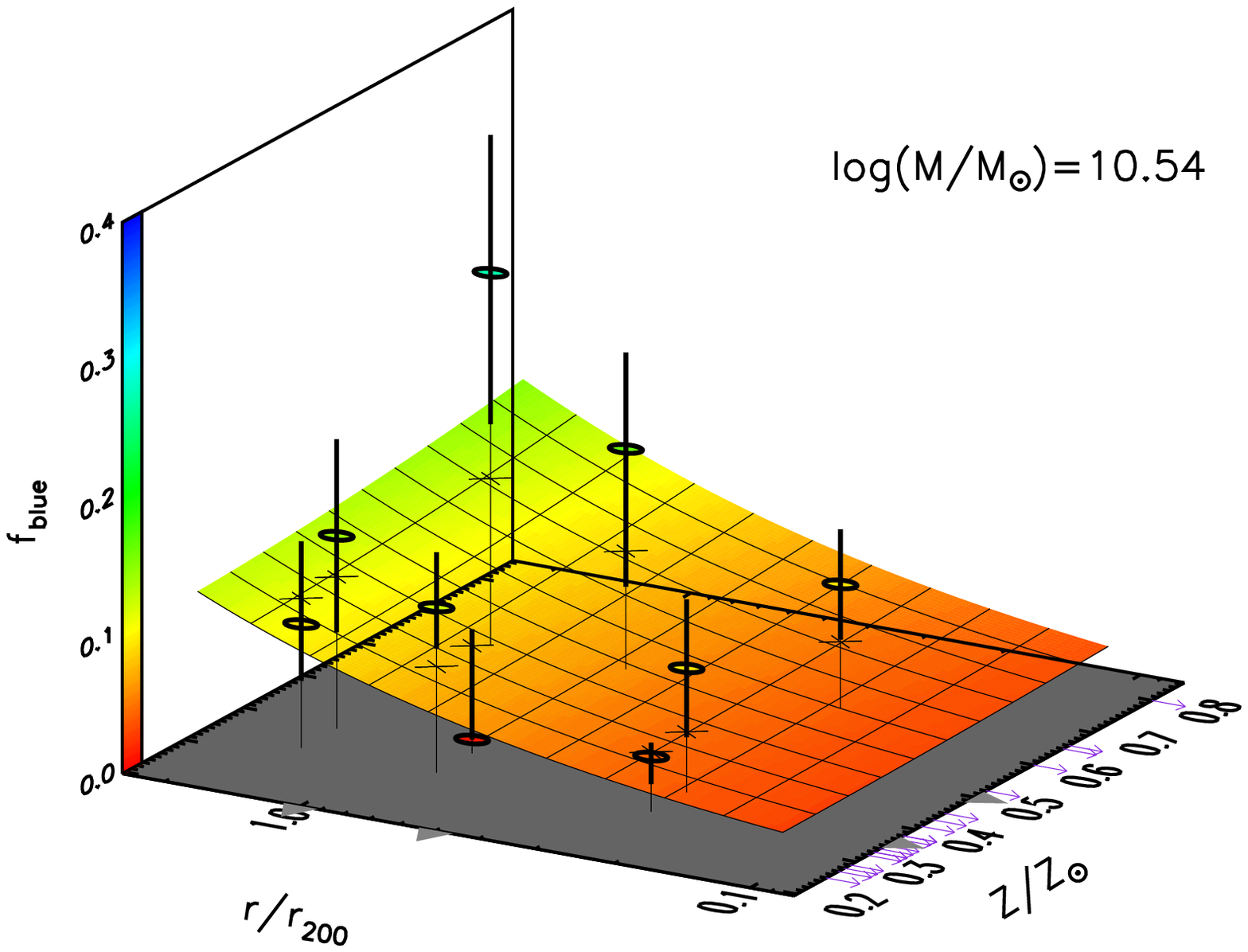} \\
	&\\
	\includegraphics[width=0.5\linewidth]{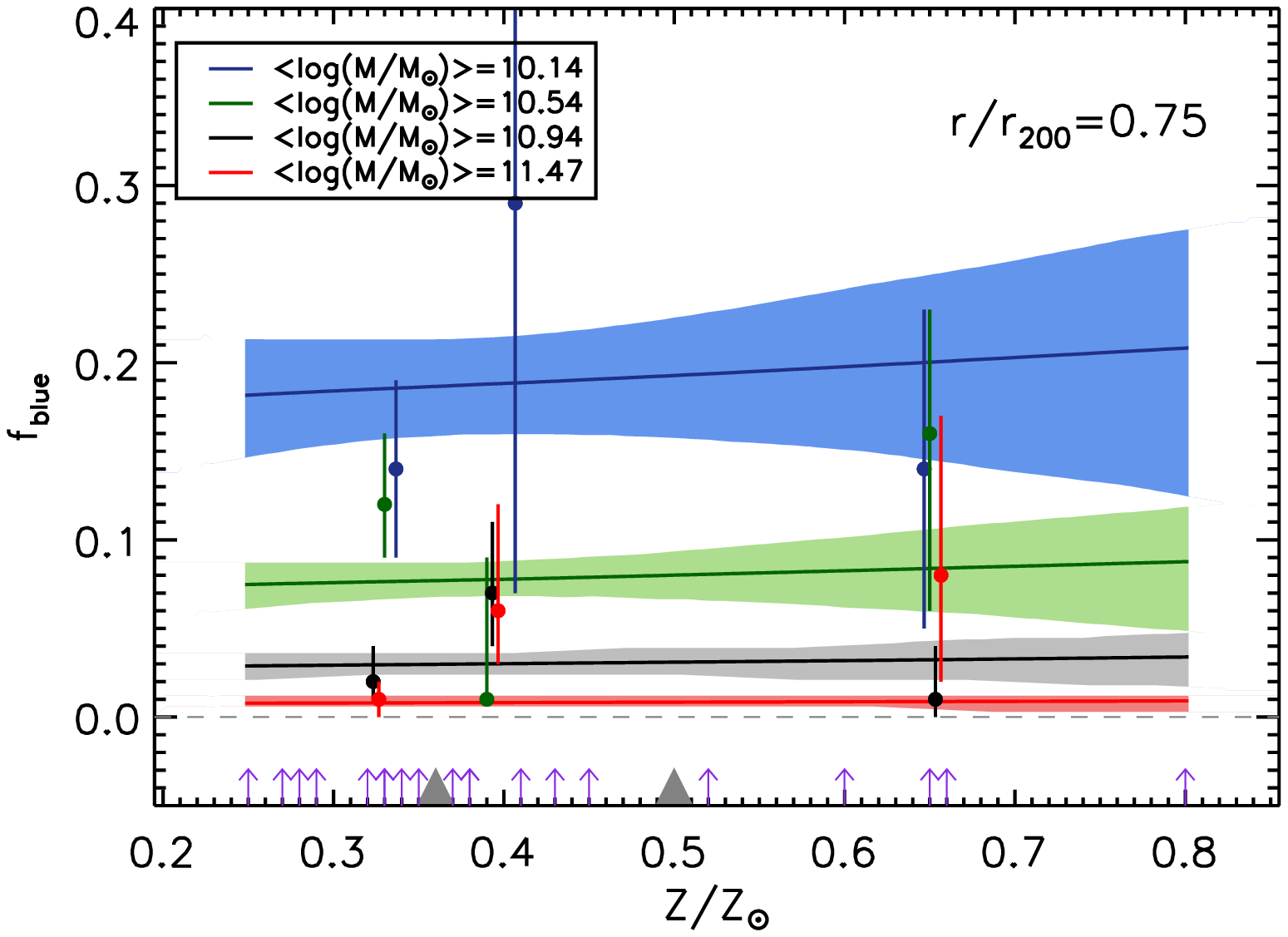} &
	\includegraphics[width=0.5\linewidth]{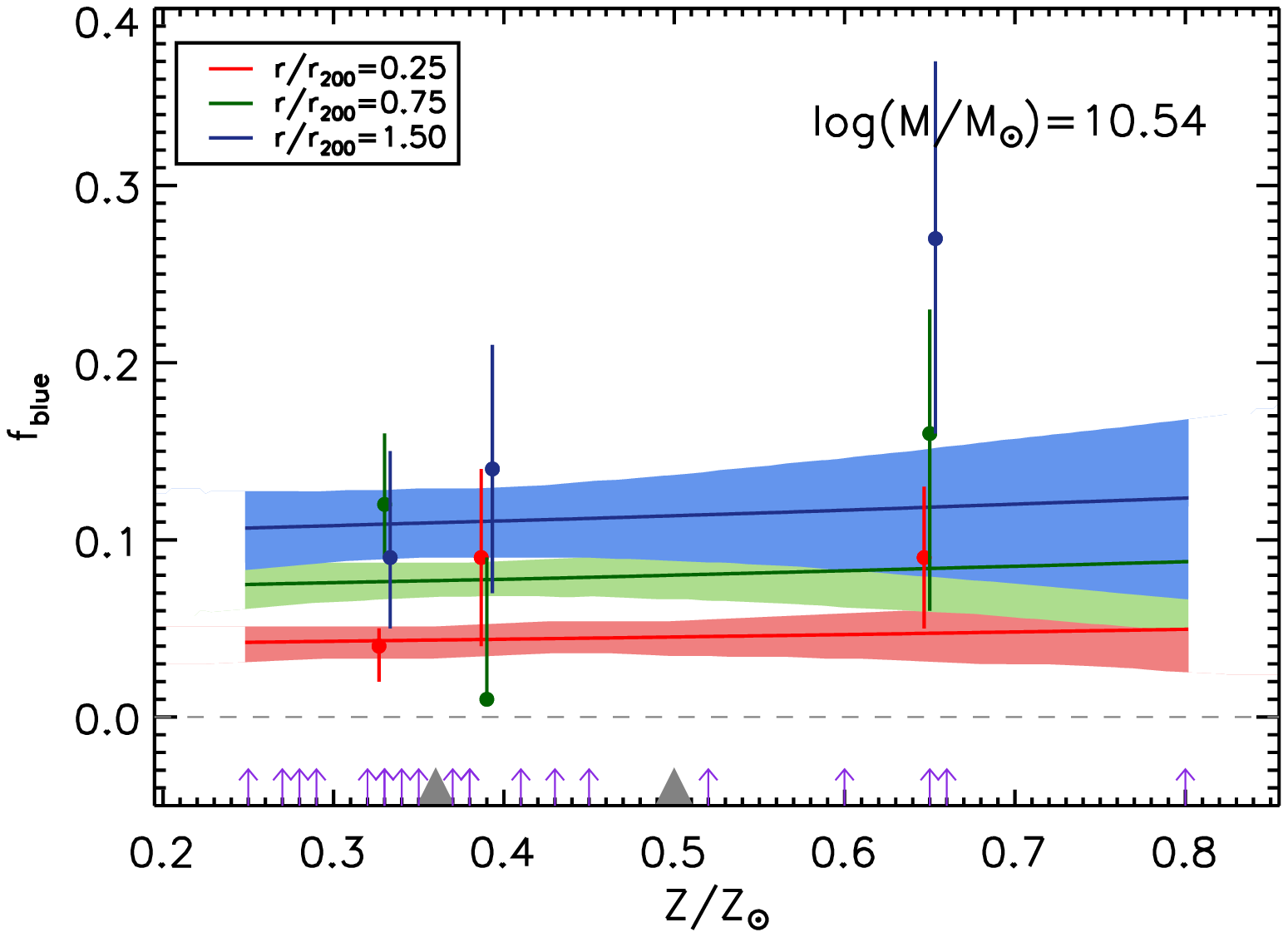}\\
\end{tabular}
\caption{
Two-dimensional slice/view of the six-dimensional parameter space describing the dependence of the quenching rate on galaxy and cluster properties.
The figures consider the following quantities: cluster iron abundance $Z/Z_\odot$, galaxy stellar mass $M$, and cluster-centric distance $r/r_{200}$.
The fit is performed on the 204 individual $f_{\rm{blue}}$ measurements.
Symbols are similar to those in Figure \ref{fig:fblue_lgM200}.
\label{fig:fblue_Zo}
}
\end{figure*}

We present the dependence of $f_{\rm{blue}}$ on cluster iron abundance in Figure \ref{fig:fblue_Zo}.
Our cluster metallicities are estimated through the spectral fit of X-ray data \citep{hudson10}.
Our fit provides a slope of $\gamma_{Z}=0.4^{+1.0}_{-1.0}$, hence indicating no significant dependence.
This implies that clusters having a larger $f_{\rm{blue}}$, at a fixed $\{$log$(M/M_\odot)$, $r/r_{200}\}$, do not necessarily have a higher iron abundance.
This result disfavours the scenario where the cluster iron abundance is rapidly increased by the addition of the metals currently produced within star-forming galaxies by Type II supernovae, which are expected to be concomitant to the star formation: either those metals are scattered into the intracluster medium on long time-scales or their contribution to the cluster iron abundance is negligible.

% FIGURE: FBLUE = f(CLUSTER METALLICITY)

% lgCCT
\subsection{Cluster central cooling time log($CCT$)}
Selecting clusters in X-ray for studies on quenching ensures a cluster selection independent of the studied quantity, contrary to an optical cluster selection.
However, an X-ray selection might be biased against/towards cool-core clusters.
The presence of a cool-core in a cluster indicates a relaxed dynamical state, as any merger event will destroy it.
It it thus interesting to study the existence of a link between the presence of a cool-core and a star-formation activity possibly induced by a merging event.

\begin{figure*}
\begin{tabular}{@{\extracolsep{5pt}} cc}
	\includegraphics[width=0.5\linewidth]{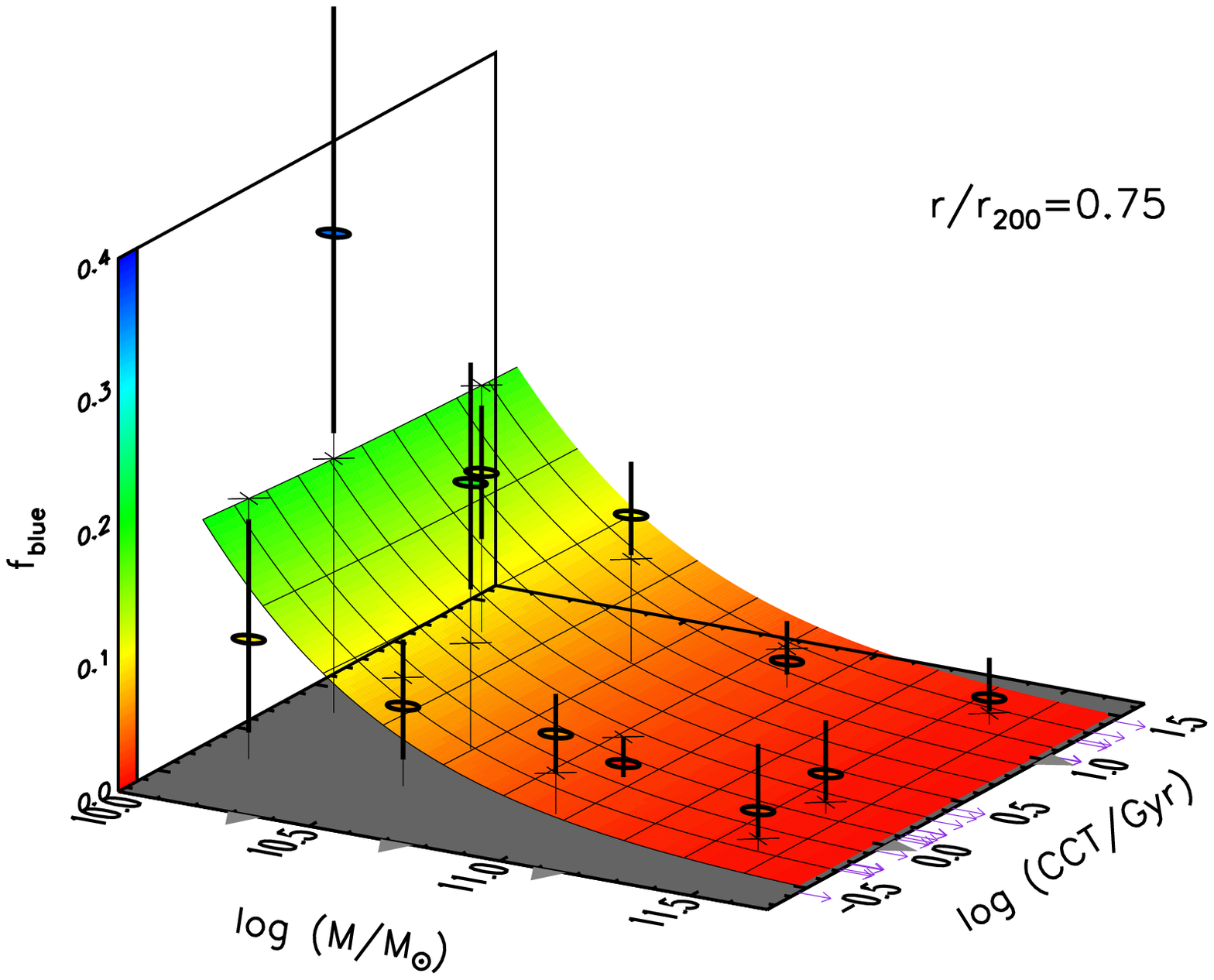} &
	\includegraphics[width=0.5\linewidth]{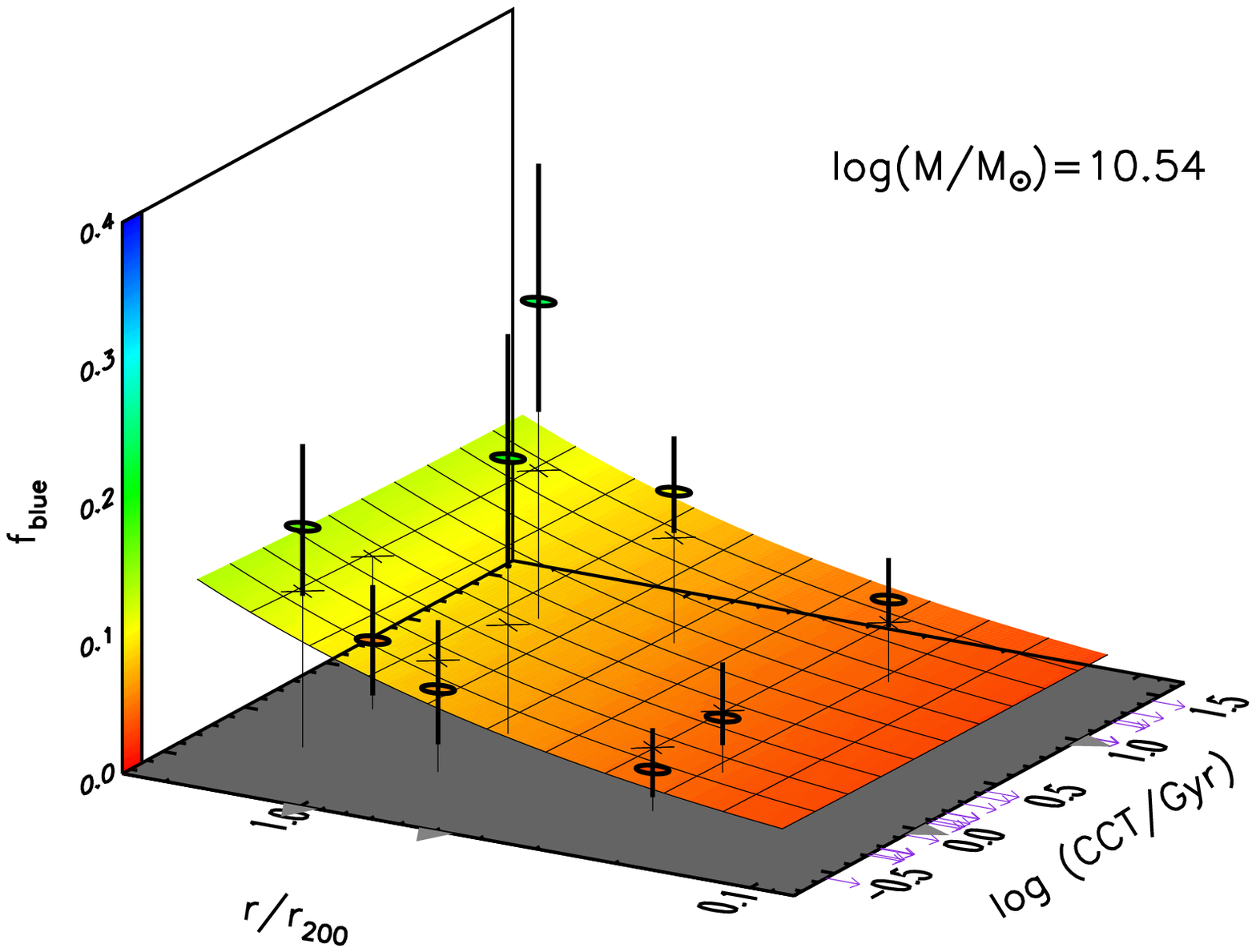} \\
	&\\
	\includegraphics[width=0.5\linewidth]{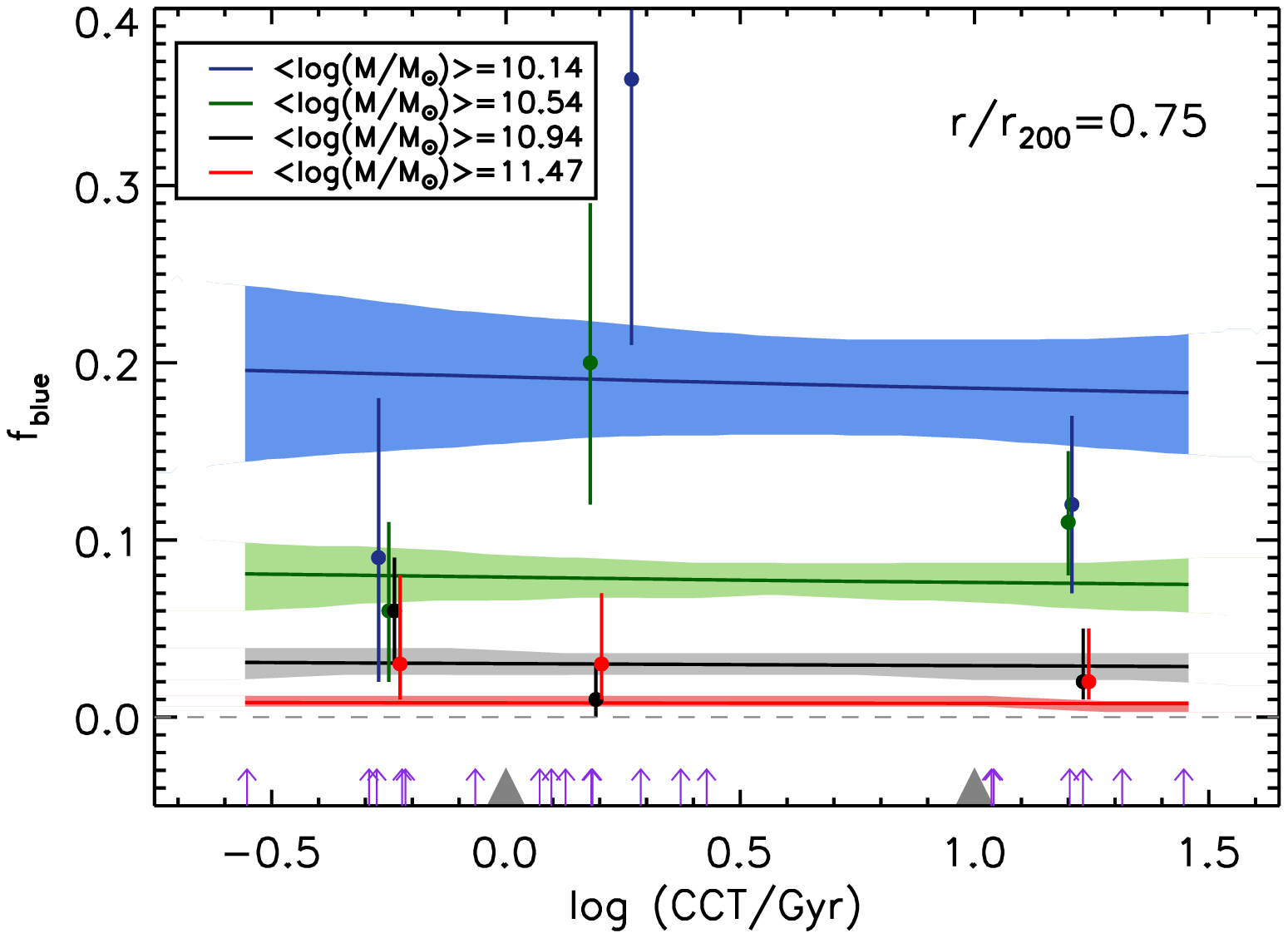} &
	\includegraphics[width=0.5\linewidth]{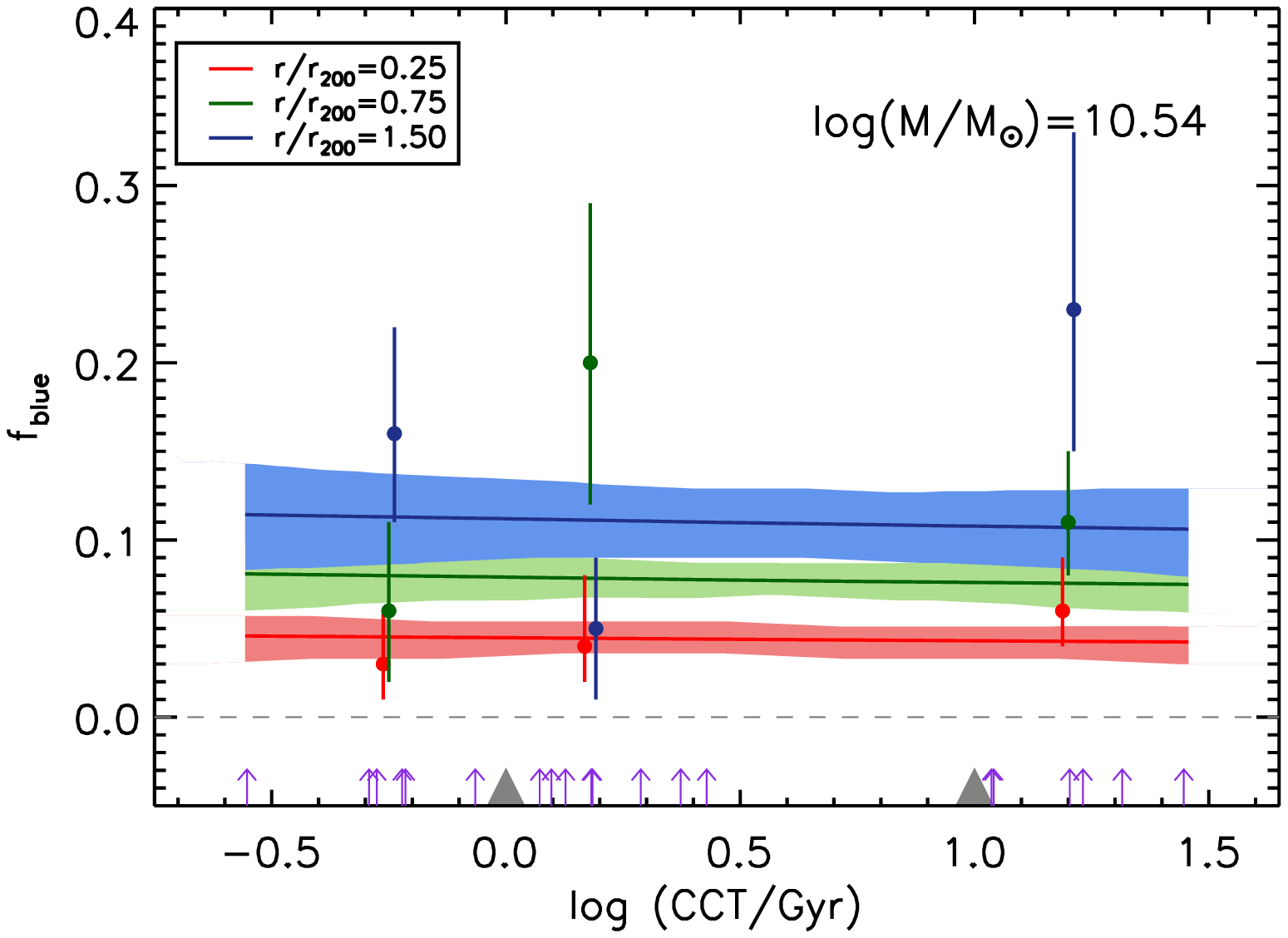}\\
\end{tabular}
\caption{
Two-dimensional slice/view of the six-dimensional parameter space describing the dependence of the quenching rate on galaxy and cluster properties.
The figures consider the following quantities: cluster central cooling time log$(CCT/Gyr)$, galaxy stellar mass $M$, and cluster-centric distance $r/r_{200}$.
The fit is performed on the 204 individual $f_{\rm{blue}}$ measurements.
Symbols are similar to those in Figure \ref{fig:fblue_lgM200}.
\label{fig:fblue_lgCCT}
}
\end{figure*}

We present the dependence of $f_{\rm{blue}}$ on cluster central cooling time log($CCT$) in Figure \ref{fig:fblue_lgCCT}.
We do not observe a dependence of $f_{\rm{blue}}$ with log($CCT$) (slope $\gamma_{CCT} = -0.1_{-0.2}^{+0.2}$).
Our result implies that clusters with evidence of a recent merger event (log($CCT/Gyr) \gtrsim 1$) do not present an increase in their fraction of blue galaxies.

In addition, this result \textit{a posteriori} validates that an X-ray cluster selection, which may be biased towards cool-core clusters due to their brighter core in X-ray emission, will not be biased regarding $f_{\rm{blue}}$.

% FIGURE: FBLUE = f(CLUSTER CCT)

% (logM,r) fit
\section{Hidden dependences and simplest model \label{sec:logMrfit}}

% Hidden dependences
\subsection{Hidden dependences}
In the previous section, we tested whether the quenching rate depends on individual halo properties (mass, richness, iron abundance and central cooling time) after accounting for the well-known dependence on galaxy stellar mass and cluster-centric distance: we found no evidence of an additional dependence, and we put a firm upper limit of 3\% to the residual effect that halos may have on the probability  that a galaxy is quenched. 
However, it might be that the lack of detection of a relation is the result of two (or more) opposite dependencies hiding each other, for example, because of parameter collinarities (e.g. a quenching rate increasing with mass and decreasing with richness will not show up, because richness and mass are tightly correlated in the sample).

For this reason, we fitted the data with a model allowing for all these four dependences at once.
In practice, we allowed four $\gamma_Q$ terms (one per halo property) in Eq. (\ref{eq:model}).
The 68\% interval of the found $\gamma_Q$'s all include zero, ruling out our hypothesis of multiple dependencies hiding each other.

% Simplest model
\subsection{Simplest model}
Given that there is no significant dependence of $f_{\rm{blue}}$ with any of the tested cluster properties, the data should be well fitted by a model including only a dependence on the galaxy stellar mass $M$ and on the cluster-centric distance $r/r_{200}$ (i.e. assuming $\gamma_Q=0$ in Eq. (\ref{eq:model})).

\begin{figure*}
\begin{tabular}{c c}
\includegraphics[width=0.5\linewidth]{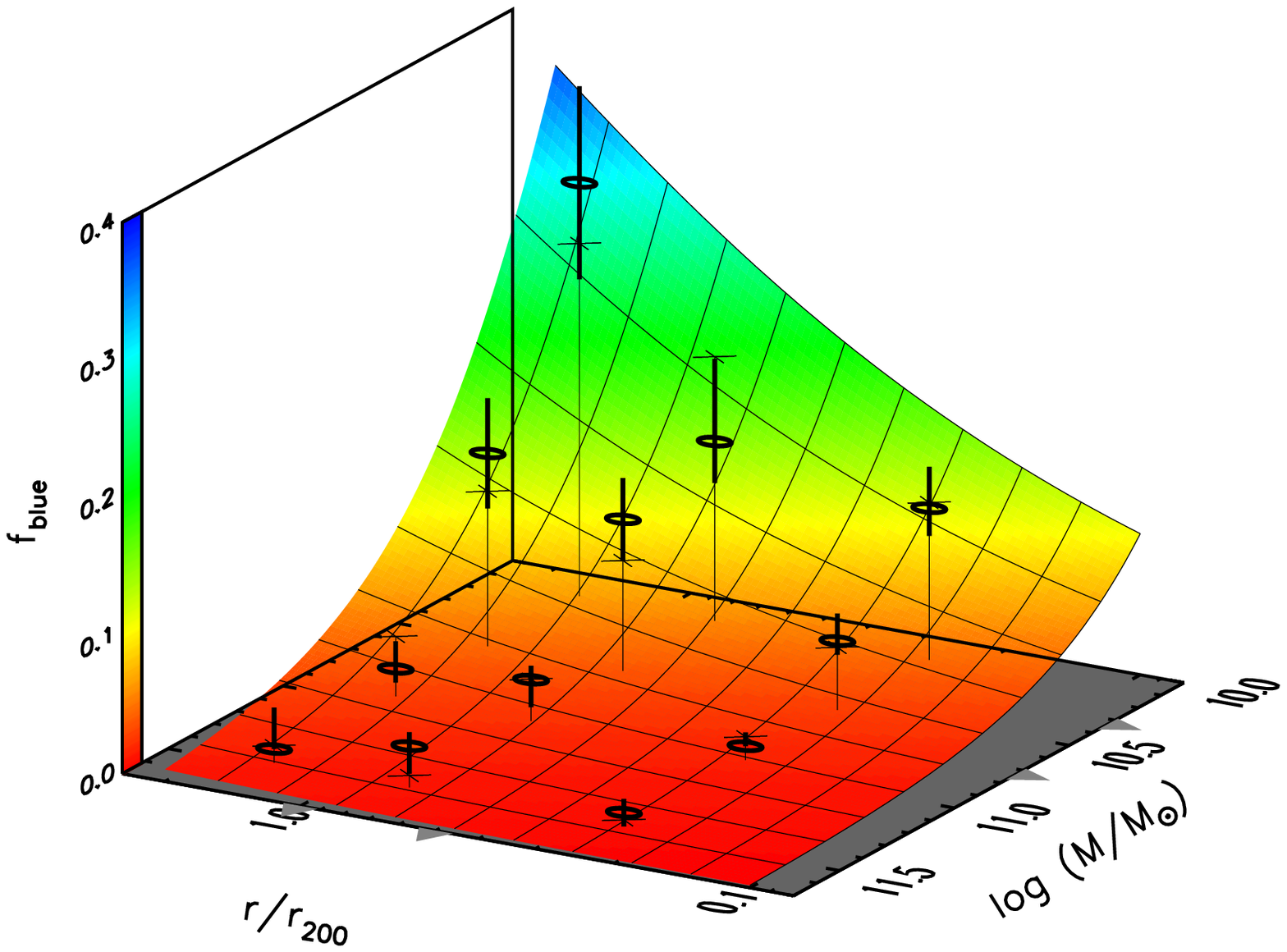} &
\includegraphics[width=0.5\linewidth]{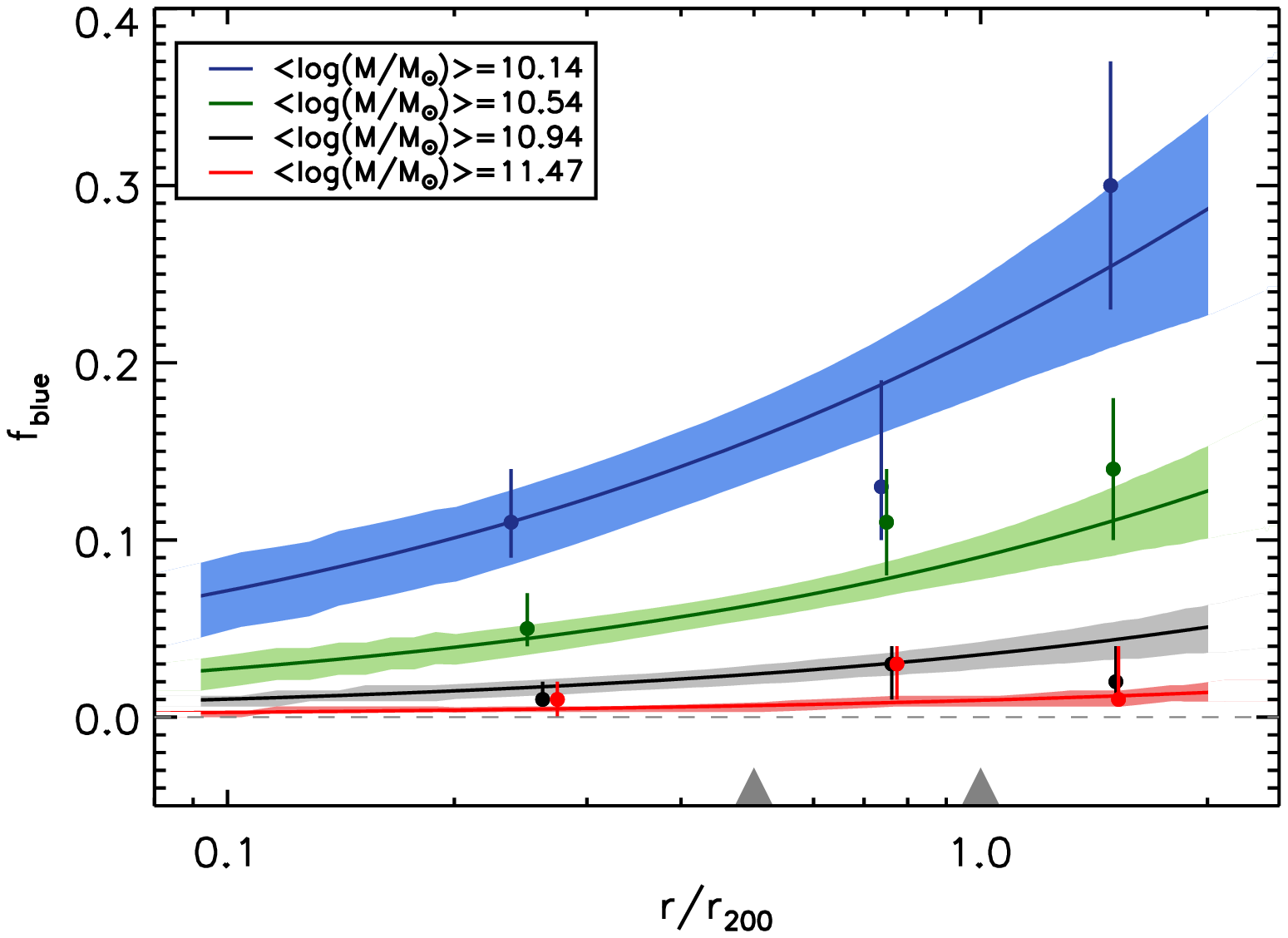}\\
\end{tabular}
\caption{
Dependence of $f_{\rm{blue}}$ on galaxy stellar mass $M$ and cluster-centric distance $r/r_{200}$.
Here the model includes only a dependence on $M$ and $r/r_{200}$ (i.e., as Eq. (\ref{eq:model}), without the $Q$-related term).
\textit{Left panel}: Thick circles with error bars represent the stacked data and the 68\% confidence interval.
The colour-coded surface represents the fitted model (mean value of the posterior distribution at each [x,y] locus).
The model value of $f_{\rm{blue}}$ at the [x,y] locus of each stacked data is symbolised by a cross.
\textit{Right panel}: Data points with error bars represent the stacked data (slightly shifted along the x-axis for clarity).
Solid lines and shaded areas represent the posterior mean and error of the model of Eq. (\ref{eq:model}).
The fit is performed on the 204 individual $f_{\rm{blue}}$ measurements.
\label{fig:lowz_stack}}
\end{figure*}

Indeed, such a model provides a convincing fit of the data, with the obtained parameters being unchanged ($A_0=4.2^{+0.3}_{-0.3}$, $\alpha=1.2^{+0.5}_{-0.3}$,  and $\beta=-2.5^{+0.4}_{-0.4}$).
Figure \ref{fig:lowz_stack}, displaying the fitted model along with the stacked data, allows us to visualise the dependence of $f_{\rm{blue}}$ on $M$ and $r/r_{200}$.
We observe that, for massive (log($M/M_\odot)>11$) galaxies, $f_{\rm{blue}}$  is very low ($<0.1$) and almost independent of $r/r_{200}$ and that, when going to smaller galaxy stellar masses, the dependence on $r/r_{200}$ increases.

Besides, as mentioned in Sect. \ref{sec:results_fbluefit}, the criterion for choosing the model parametrisation was only simplicity \citep[see also][]{raichoor12b}.
We now have data of good enough quality to check whether the model is flexible enough to fit the data (by stacking the data only in $M$ and $r/r_{200}$ bins, we have 12-20 individual points per bin).
We observe that the analytical choice of our model is fully able to fit the data, thus justifying \textit{a posteriori} the chosen analytical form of the model.\\

% FIGURE: SDSS - STACKED

%@@@@@@@@@@@@@@@@@@@@@@@@@@@@@@@@@@@@@@@@@@@
% DISCUSSION AND CONCLUSIONS
%@@@@@@@@@@@@@@@@@@@@@@@@@@@@@@@@@@@@@@@@@@@

\section{Discussion and conclusions \label{sec:conclusion}}

% method summary
We studied how the quenching rate in galaxies with log$(M/M_\odot)>10$ and within two virial radii depends on cluster properties.
To that end, we used a low-redshift cluster sample ($0.02 \le z \le0.1$ and $13.9 \le \textnormal{log}(M_{200}/M_\odot) \le 15.2$).
This cluster sample is selected independent of the quantity being studied (in X-ray) and has homogeneous, well defined properties (mass, richness, iron abundance, and central cooling time).
We used the fraction of blue galaxies $f_{\rm{blue}}$ as a proxy for tracing the fraction of quenched galaxies.
We carefully estimated $f_{\rm{blue}}$ for each cluster, as a function of galaxy stellar mass and cluster-centric distance, accounting for background contamination.
For each cluster property, we then fitted a model to our data, looking for dependency on galaxy stellar mass, cluster-centric distance, and the tested cluster property.

% r and logM quenching
We found that $f_{\rm{blue}}$ depends at the same time on the galaxy stellar mass and on the cluster-centric distance, and that the two associated quenching modes, mass and environmental quenching, are separable.
This result, in agreement with previous studies \citep[e.g.][]{peng10,raichoor12b}, illustrates the requirement to lead such analysis as a function of galaxy stellar mass and environment at the same time.

% result
On top of this well-known dependence of the quenched fraction on galaxy stellar mass and environment, our analysis does not find any significant dependence of the quenching rate on the cluster properties, other than that induced by the $M_{200}-r_{200}$ relation: for all the four tested quantities, a null slope is (almost) included in the 68\% confidence interval.
Our tested cluster properties probe 
halo mass ($13.9 \le \textnormal{log} (M_{200}/M_\odot) \le 15.2$), 
richness ($1.0 \le \textnormal{log}(N_{200}) \le 2.2$),
metal content ($0.25 \le Z/Z_\odot \le 0.80$),
and relaxation state ($-0.6 \le \textnormal{log} (CCT/Gyr) \le 1.4$).
Hence this study links the quenching rate to different aspects of the cluster history and increases our knowledge of possible selection effects when building cluster samples.

% no starburst in merging clusters
Our finding that $f_{\rm{blue}}$ is independent of the central cooling time is evidence of a lack of starbursts in clusters observed during, or just after, a merging event.
Clusters with large central cooling time  ($0.9 < \textnormal{log}(CCT/Gyr)$) are expected to be in the course of, or just
passed, a merging episode: in fact, five out of the six non-cool core clusters in our sample show sign of merging activity \citep{hudson10}.
Our data and model do not point towards an increase of $f_{\rm{blue}}$ for those clusters.
Hence, we do not observe evidence for an increase number of starburst galaxies in merging clusters.
We must however stress as our study is based on photometry, we lack sensitivity on starburst episodes/features only perceptible with spectroscopy.

% X-ray selection unbiased towards fblue
This independence of $f_{\rm{blue}}$ on the central cooling time validates \textit{a posteriori} that an X-ray cluster selection will not be biased regarding $f_{\rm{blue}}$.

% cluster mass
Our result that $f_{\rm{blue}}$ does not depend on the cluster mass (or richness), once the cluster-centric distance is normalised by $r_{200}$, indicates that mass-dependent selection effects are negligible for X-ray selected clusters more massive than $10^{14} M_\odot$.
This result enables the enlargement of future cluster samples for similar studies, as it relaxes the constraint of comparing local clusters to their likely ancestors.
Further studies might show whether if removing either of the two limits (X-ray selection or mass threshold) is safe.
While we found no trend with mass and richness, this does not guarantee that this holds for clusters of lower mass or richness, or selected in a different way.

% metallicity
We did not find any dependence of $f_{\rm{blue}}$ on the cluster iron abundance.
This result disfavours the scenario where the cluster iron abundance is rapidly increased by the addition of the metals produced within cluster galaxies by Type II supernovae, which are expected to be concomitant to the star formation: either those metals are scattered into the intracluster medium on long time-scales, or their contribution to the cluster iron abundance is negligible.\\

% conclusion
In this study, our requirement of a minimal signal to noise ratio of 5 in $u-r$ in the SDSS photometry prevented us from exploring the behaviour of such relations at low galaxy stellar masses (log($M/M_\odot) < 10$).
It would be promising to extend the analysis to such low galaxy stellar masses with deeper data, as it would allow us to extensively probe a galaxy stellar mass regime where galaxies are observed to have more star-formation \citep[e.g.][]{kauffmann03a,kauffmann04} , hence where quenching efficiency is expected to be more visible.

%@@@@@@@@@@@@@@@@@@@@@@@@@@@@@@@@@@@@@@@@@@@
% BIBLIOGRAPHY
%@@@@@@@@@@@@@@@@@@@@@@@@@@@@@@@@@@@@@@@@@@@

%@@@@@@@@@@@@@@@@@@@@@@@@@@@@@@@@@@@@@@@@@@@
% ACKNOWLEDGMENTS
%@@@@@@@@@@@@@@@@@@@@@@@@@@@@@@@@@@@@@@@@@@@

\begin{acknowledgements}

We acknowledge financial contribution from the agreement ASI-INAF I/009/10/0 and from Osservatorio Astronomico di Brera.

Based on data from SDSS-III (full text acknowledgement is at \url{http://www.sdss3.org/collaboration/boiler-plate.php}).

\end{acknowledgements}

%@@@@@@@@@@@@@@@@@@@@@@@@@@@@@@@@@@@@@@@@@@@
% APPENDIX
%@@@@@@@@@@@@@@@@@@@@@@@@@@@@@@@@@@@@@@@@@@@

\appendix

\section{Additional material}

\begin{figure*}
	\includegraphics[width=\linewidth]{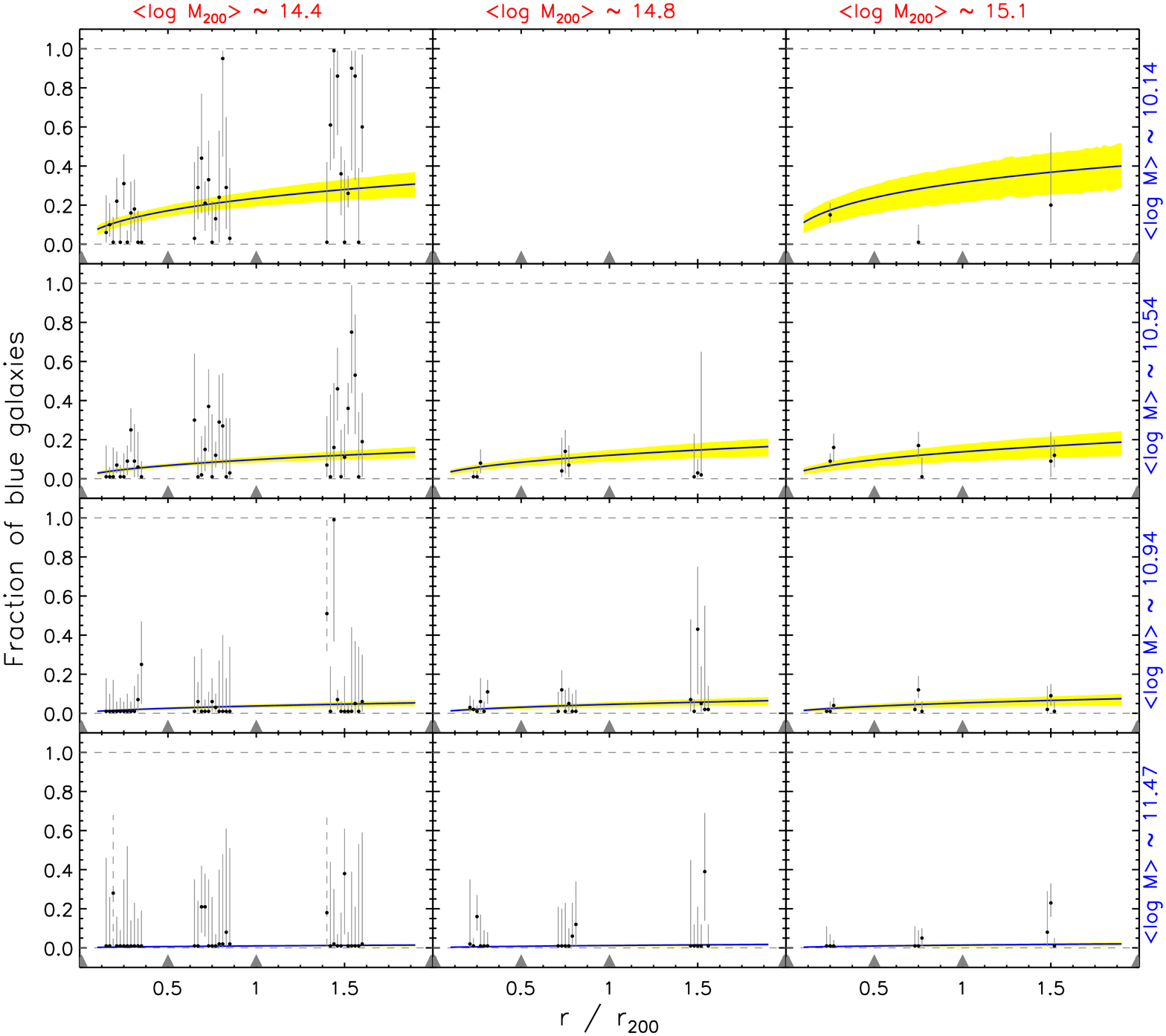}\\
\caption{
We display $f_{\rm{blue}}$ for individual clusters as a function of cluster-centric distance ($r/r_{200}$) for different bins of cluster mass log($M_{200}/M_\odot$) (increasing right-ward) and galaxy stellar mass (increasing down-ward).
Error bars mark the shortest interval including 68\% of the posterior values, and are plotted as a dashed line when this interval is larger than 0.66, indicating that $f_{\rm{blue}}$ is very poorly constrained.
Yellow shaded areas represent the posterior mean and error of the model of Eq. (\ref{eq:model}), fitting the 204 individual $f_{\rm{blue}}$ measurements.
Radial cluster-centric bins are indicated by grey filled triangles on the x-axis and horizontal grey dashed lines limit the whole $f_{\rm{blue}}$ range.
\label{fig:fblue_lgM200_indiv}
}
\end{figure*}

We display in Figure \ref{fig:fblue_lgM200_indiv} the 204 measured $f_{\rm{blue}}$ values, as a function of galaxy stellar mass $M$, cluster-centric distance $r/r_{200}$, for our three cluster mass bins along with our fitted model.
We notice that there are few points that are more than $\sim$2$\sigma$ away from the model, which is expected when there are 204 measurements.\\

% FIGURE: indiv. fblue (lgM200)

\begin{figure*}
	\includegraphics[width=\linewidth]{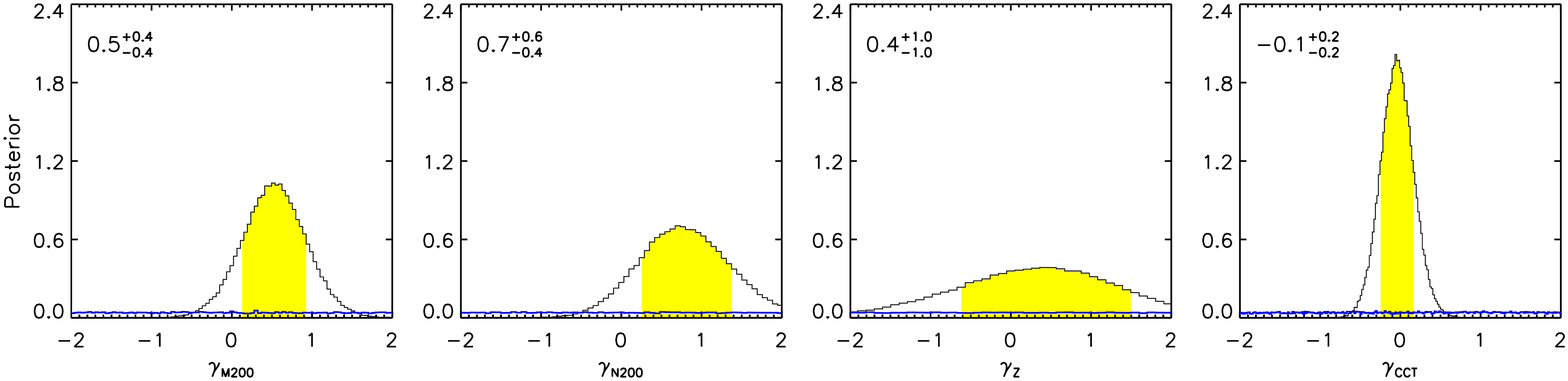}\\
\caption{
Computed posterior for the $\gamma_Q$ coefficient of Eq.(\ref{eq:model}).
From left to right, $Q$ stands for: cluster mass $\textnormal{log}(M_{200}/M_\odot)$, cluster richness $\textnormal{log}(N_{200})$, cluster iron abundance $Z/Z_\odot$, and cluster central cooling time $\textnormal{log}(CCT/\textnormal{Gyr})$.
The posterior is the black histogram and the prior the blue histogram, close to the x-axis).
The yellow shaded area represents the 68\% interval.
For each quantity $Q$, we report the point estimate and the 68\% interval.
\label{fig:gamma_post}
}
\end{figure*}

We display in Figure \ref{fig:gamma_post} the computed posterior for the $\gamma_Q$ coefficient of Eq.(\ref{eq:model}) for $Q$ in: cluster mass $\textnormal{log}(M_{200}/M_\odot)$, cluster richness $\textnormal{log}(N_{200})$, cluster iron abundance $Z/Z_\odot$, and cluster central cooling time $\textnormal{log}(CCT/\textnormal{Gyr})$.
For each quantity $Q$, we report the point estimate and the 68\% confidence interval.

% FIGURE: gamma posterior

\end{document}